\documentclass[10pt,preprint]{aastex}

\def\lapprox{\mathrel{\hbox{\rlap{\hbox{\lower4pt\hbox{$\sim$}}}\hbox{$<$}}}}
\def\gapprox{\mathrel{\hbox{\rlap{\hbox{\lower4pt\hbox{$\sim$}}}\hbox{$>$}}}}

\newcommand{\be}{\begin{equation}}
\newcommand{\ee}{\end{equation}}

\begin{document}
\shorttitle{CME Initiation by Flux Emergence}
\shortauthors{Leake et al.}

\title{Tests of Dynamical Flux Emergence as a Mechanism for CME Initiation}

\author{James E. Leake}
\affil{U.S. Naval Research Lab 4555 Overlook Ave., SW Washington, DC 20375. \\
College of Science, George Mason University, 4400 University Drive, Fairfax, Virginia 22030. \\
james.leake.ctr.uk@nrl.navy.mil}

\author{Mark G. Linton}
\affil{U.S. Naval Research Lab 4555 Overlook Ave., SW Washington, DC 20375.\\
mark.linton@nrl.navy.mil}

\and

\author{Spiro K. Antiochos}
\affil{NASA Goddard Space Flight Center,
Heliophysics Division,
NASA/GSFC,
Greenbelt, MD 20771.\\
spiro.anticohos@nasa.gov}

\begin{abstract}
Current coronal mass ejection (CME) models set their lower boundary to be in the lower corona. They do not calculate accurately the  transfer of free magnetic energy from the convection zone to the magnetically dominated corona because they model the
effects of flux emergence using kinematic boundary conditions or simply assume the appearance of flux at these heights. We test the importance of including dynamical flux emergence in CME modeling by simulating, in 2.5D,  the emergence of sub-surface flux tubes into different coronal magnetic field configurations. We investigate how much free magnetic energy, in the form of shear magnetic field, is transported from the convection zone to the corona, and whether dynamical flux emergence can drive CMEs. We find that multiple coronal flux ropes can be formed during flux emergence, and although they carry some shear field into the corona, the majority of shear field is confined to the lower atmosphere. Less than 10\% of the magnetic energy in the corona is in the shear field, and this, combined with the fact that the coronal flux ropes bring up significant dense material, means that they do not erupt.  Our results have significant implications for all CME models which rely on the transfer of free magnetic energy from the lower atmosphere into the corona but which do not explicitly model this transfer. Such studies of flux emergence and CMEs are timely, as we have new capabilities to observe this with Hinode and SDO, and therefore to test the models against observations.           
\end{abstract}
\section{INTRODUCTION}

\subsection{CME modeling}
Coronal mass ejections (CMEs) and eruptive flares are the most
energetic manifestations of solar activity, with a typical CME
accelerating $10^{16}$ g of coronal plasma to speeds sometimes
exceeding 1000 km/s
\citep{1976SoPh...48..389G,1977ASSL...71....3H}. It is now generally
accepted that these giant eruptions of solar plasma and field are due
to the explosive release of magnetic energy stored in the corona prior
to the event. Furthermore, the
free energy for CMEs/flares is believed to be stored in the strongly
sheared flux of filament channels (see, e.g, reviews by
\citet{2000JGR...10523153F,Klimchuk2001,LintonM2009}).  
This flux is {\it sheared} in
that that the field lines are mainly parallel to the photospheric
neutral line, rather than perpendicular as would be expected for a
potential field. All CMEs/flares are associated with filament channels
and these are the only locations in the corona where the magnetic
field exhibits strong non-potentiality. The basic picture of a CME is that,
prior to the event, there is a force balance in the corona between the
upward magnetic pressure of the sheared filament channel field and the
downward tension of overlying unsheared field. As the shear builds up
or the overlying field evolves, this force balance eventually disrupts
producing an explosive outward expansion of the filament channel flux
and some of the overlying flux.

It is evident from this discussion that the two central issues in
understanding CMEs/eruptive flares are the process by which magnetic
shear forms in filament channels, and the mechanism that disrupts the
force balance. These two issues have been at the heart of CME and
flare research for decades and are the features that distinguish the
various CME models. The first issue of the shear formation
process is especially critical, because the disruption mechanism is
almost certain to depend on the topology of the filament channel
field, which must be determined by the formation process.  For
example, some models invoke kink-like instabilities, which
requires the formation of magnetic twist as well as shear in the
filament channel field. We emphasize, however, that all the models and
all the observations agree that the bulk of the magnetic free energy
is in the form of shear. Note also that the issue of shear
formation has broad implication for a number of important solar
physics problems, in particular the origin of prominences and
filaments.

There are only two general processes that can produce the observed
shear in filament channels: either the field emerges from below in a
non-potential, sheared state, or the photospheric motions shear the
field after it emerges into the corona. Of course, both processes must
be present to some extent in the real corona. The latter process is
likely to dominate in non-active region filament channels, such as
those associated with high-latitude quiescent filaments, because these
form well after the flux has emerged. On the other hand, in active
regions, especially in the strong complex regions that are the source
of fast CMEs, the filament channels form with the active region and,
hence, the shear must emerge with the flux. Given that the shear
formation is the fundamental driver of the eruption, it is clear that
flux emergence must be explicitly included in any model for fast CMEs
in order for the model to be physically rigorous.

\begin{figure}
\begin{center}
\includegraphics[width=\textwidth]{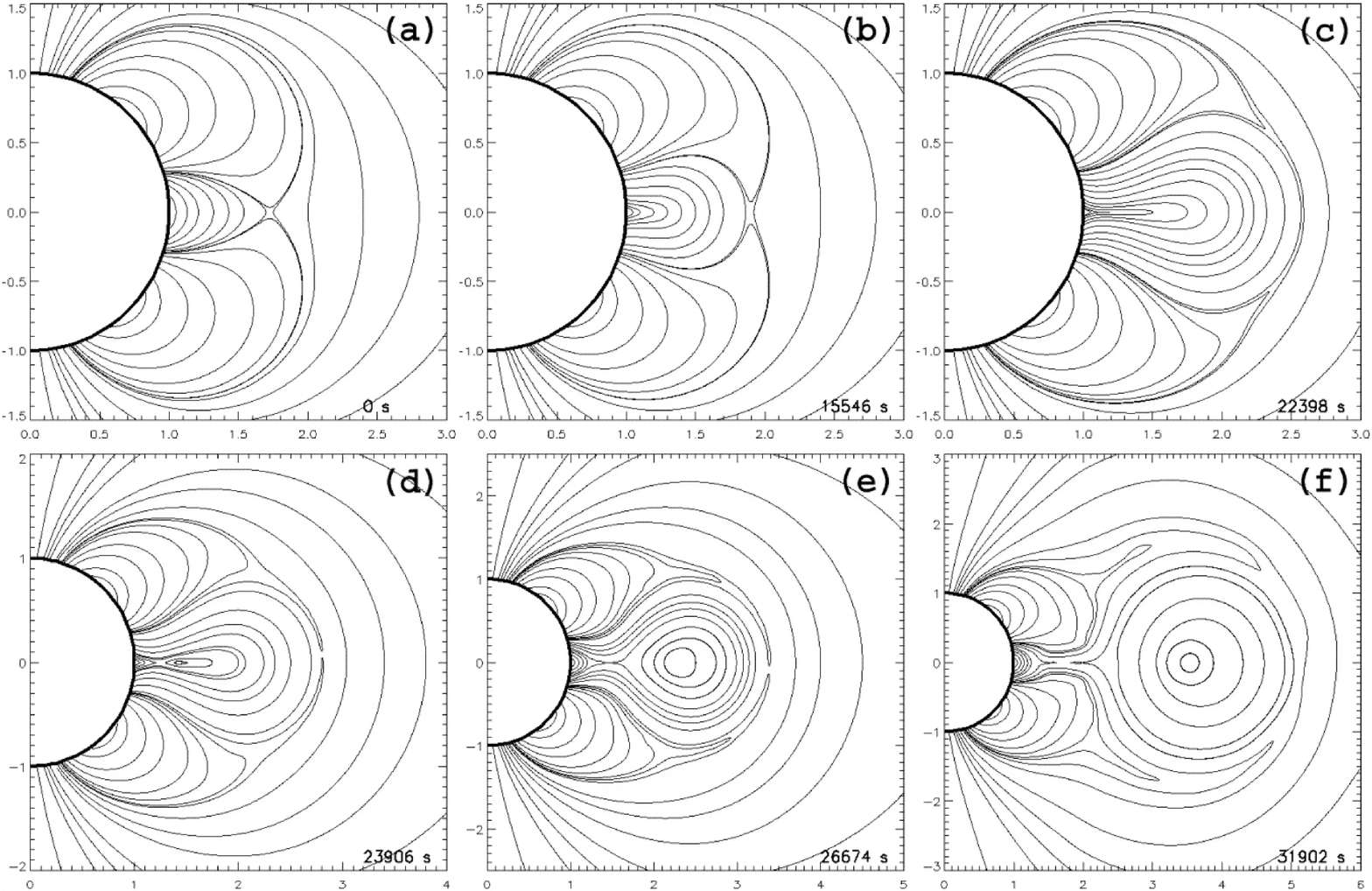}
\caption{Meridional projections of magnetic field lines throughout the magnetic breakout eruption process at six different times. Reprinted with permission from \citet{2004ApJ...617..589L}. Spatial units are solar radii.
\label{fig:lynch04_fig1}}
\end{center}
\end{figure}

\begin{figure}
\begin{center}
\includegraphics[width=\textwidth]{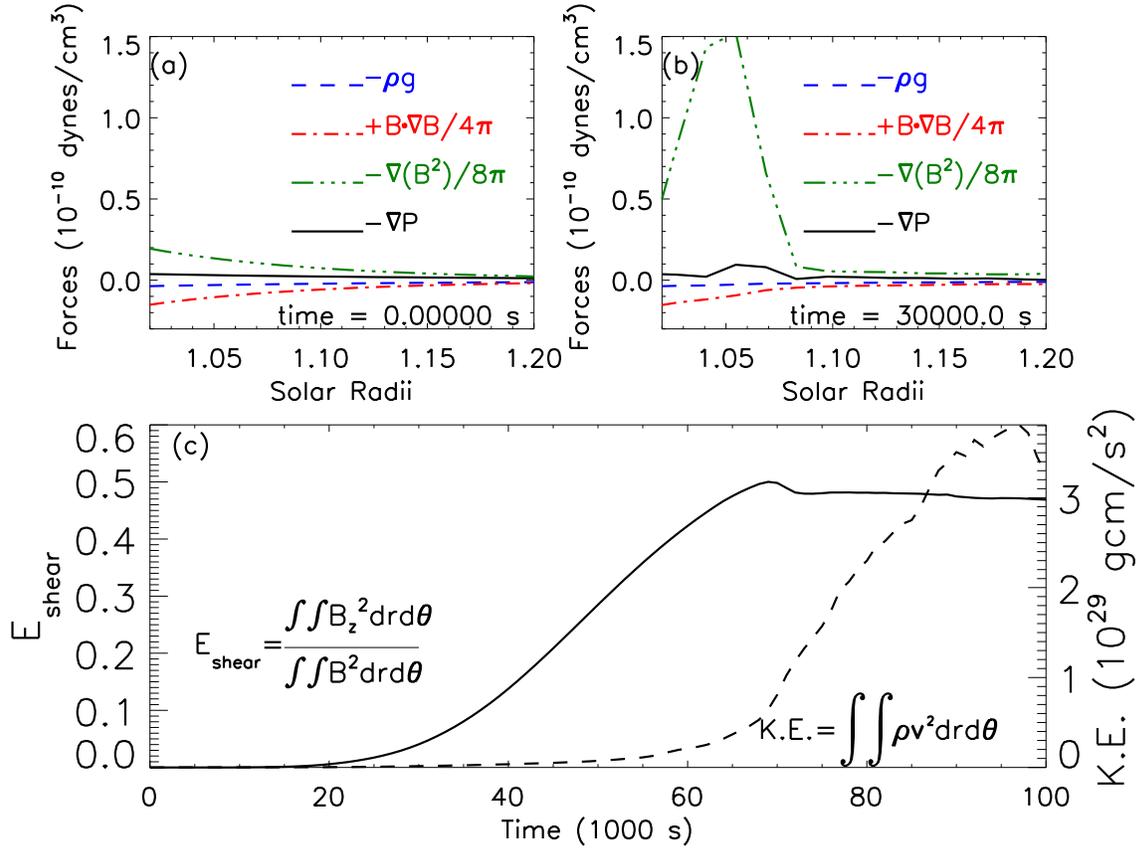}
\caption{Panels (a) and (b): Vertical forces along the latitude=$0^{\circ}$ line at the center of the inner arcade of the magnetic quadrupole in the breakout model at t=0 s, panel (a), and  t=30000 s, panel (b), after shearing of the arcade by imposed boundary flows. Panel (c): The normalized shear magnetic energy (solid line) and 2D integral of kinetic energy density
(dashed line) in the system during this shearing period. 
\label{fig:BO_energy_forces}}
\end{center}
\end{figure}

One such model, which generates fast CMEs from sheared filaments, but which should be modified to include the creation of these sheared filaments via flux emergence, is the magnetic breakout model \citep{1998ApJ...502L.181A,1999ApJ...510..485A}. This model exploits the finite resistivity in the solar atmosphere to disrupt the pre-CME force balance by generating reconnection external to the filament channel.    This reconnection reconfigures the magnetic field and allows the eruption of the highly sheared structure. The breakout model thus requires a multipolar flux distribution with a separatrix where reconnection can occur. Numerical simulations have investigated this
model in both 2.5D, where the domain is axisymmetric and  two dimensional but where all three components of vector variables evolve \citep{2004ApJ...617..589L,2004ApJ...614.1028M,2005ApJ...628.1031D} and 3D, where the domain is three dimensional \citep{2008ApJ...683.1192L,2008ApJ...680..740D}. The basic 
model uses a current-free
quadrupole magnetic field configuration, consisting of a central 
\textit{active region} dipole embedded in an anti-parallel global dipole field, with a null point separating the two systems. This  is shown in Figure \ref{fig:lynch04_fig1}(a), taken from \citet{2004ApJ...617..589L}. Figure \ref{fig:BO_energy_forces}(a) shows, for this experiment, the vertical force balance in the central arcade along the $\textrm{latitude}=0^{\circ}$ symmetry line for the initial equilibrium. The gravitational ($\rho\mathbf{g}$) and gas pressure ($-\nabla P$) forces balance, as do the magnetic tension ($\mathbf{B}.\nabla\mathbf{B}/4\pi$) and magnetic pressure (-$\nabla{B}^{2}/8\pi$). Shear flows are applied at the 
lower boundary, within the central arcade, to generate magnetic field parallel to the active region neutral line (in our 2.5D simulations the \textit{shear field} will be the magnetic field in the ignorable direction). Figure \ref{fig:BO_energy_forces}(b) shows the vertical forces  after a short period of shearing.
 The presence of shear field exerts an 
upward magnetic pressure on the overlying field and stretches the original null point
 into a current sheet, as shown in Figures \ref{fig:lynch04_fig1}(b) and (c).  Reconnection occurs when the current sheet collapses below grid scale and overlying field is  removed by this reconnection, which allows further expansion of the central arcade. The outward expansion of the sheared arcade occurs exponentially as more and more
overlying field is reconnected. Flare-like reconnection underneath the sheared dipole creates a disconnected flux rope which can then escape, as seen in Figures \ref{fig:lynch04_fig1}(d)-(f), at speeds comparable to observations of a fast CME.

There are numerous other CME models which either assume the appearance of sheared magnetic flux at the low corona, or use driving velocities to create sheared structures from coronal equilibria. \citet{2000ApJ...545..524C} and \citet{2006A&A...459..927D} simulated a 2.5D CME model where a pre-existing coronal flux rope, constrained by overlying field, is allowed to escape when kinematically imposed emerging flux causes reconnection in a filament channel beneath the flux rope. They also found that CME eruptions occur when kinematic flux emergence is used to drive reconnection outside the filament channel.
The flux cancellation model \citep{2000ApJ...529L..49A,
2003ApJ...585.1073A,2007ApJ...671L.189A}  is another model based on reconnection, but this model uses both boundary imposed shearing and the cancellation of magnetic flux at the neutral line of filament 
sites to initiate a CME.  \citet{2007ApJ...668.1232F} developed a CME model where the lower corona is driven by the kinematic emergence of a sheared flux rope into a pre-existing coronal field. Here, loss of equilibrium via the kink or torus instability leads to an eruption, with the strength of the overlying field being an important constraint on the eruption. The breakout model has also been extended to cover different magnetic field configurations, and also to include a solar wind, in both 2.5D and 3D simulations \citep{2007ApJ...671L..77V,2008ApJ...689L.157Z,2009A&A...501.1123S}. Other CME models exist which do not directly rely on flux emergence, such as those based on the loss of equilibrium of a coronal flux rope 
\citep{1999A&A...351..707T,2003ApJ...588L..45R,2006PhRvL..96y5002K}, and simulations of the rise of a pre-formed coronal flux rope into the solar wind by \citet{2009JGRA..11405107Z}. Although not relying on flux emergence directly, these models do assume that the flux ropes have already been formed in the corona and reached an equilibrium with the background coronal field, a state which is most likely caused by the emergence of magnetic field from beneath the surface. 

All these models, constrained by the need to extend the simulation domain to at least a few solar radii, do not model the lower solar atmosphere.  The lower boundary of these simulations has a typical density 
of $3\times10^{-16} \ \textrm{g/cm}^3$ and a typical plasma-$\beta<0.2$ where
\be
\beta = \frac{p}{B^{2}/8\pi}.
\ee
Based on the VAL model of the solar atmosphere \citep{1981ApJS...45..635V} the photosphere, in contrast,  has a
 density of around 
$3\times10^{-7} \ \textrm{g/cm}^{3}$ and $\beta>1$. These CME initiation models simply assume the appearance of flux at the low density, low $\beta$ corona and do not self-consistently calculate a process for the flux emergence. These simulations do not, therefore, address the critical question of whether and how newly emerging, sheared magnetic flux can rise from its origins in the high $\beta$ convection zone to the low corona where it is required to drive CME models.  This is an 
important question to address now, as Hinode and SDO scientists are currently making detailed observations of flux emergence with vector magnetograms, which will allow the comparison of observations and state of the art CME models.

\subsection{Flux Emergence}
The current state of the theory of the formation of active regions suggests that dynamo actions in the solar interior create twisted, buoyant flux tubes which rise through the convection zone,  intersect the photosphere and create the observed sunspots and bipolar active regions \citep{1955ApJ...122..293P,1970SoPh...14..328Z,1979A&A....71...79S,1988ApJ...326..407P,1987ARA&A..25...83Z,1998ApJ...492..804E,2000SoPh..192..119F}. The subsequent expansion of these flux tubes into the corona has been a subject of vigorous research for the past 30 years.  The most likely driver of flux emergence into the lower corona is an ideal MHD instability driven by magnetic buoyancy (the magnetic buoyancy instability), a particular mode of which was first suggested by \citet{1989ApJ...338..471S,1989ApJ...345..584S} in 2D simulations using a crude model atmosphere and magnetic field configuration. They found that the rise of magnetic flux into the corona due to this instability matched observed rise velocities of magnetic flux. 

The process of flux emergence has been studied more recently in 3D simulations with a more detailed model atmosphere that includes a convection zone, isothermal photosphere/chromosphere, transition region, and isothermal corona, with a buoyant twisted flux tube embedded in the convection zone
\citep{2001ApJ...554L.111F,2004ApJ...610..588M,2004A&A...426.1047A,2009ApJ...697.1529F}.  The reader is directed to a comprehensive review of this subject by \citet{2008JGRA..11303S04A}. These simulations show that as the flux tube intersects the photosphere, bipolar active regions are formed which exhibit shear flows along their neutral line. As the tube continues to emerge into the corona, sigmoid-like structures are formed consistent with observations, with magnetic field being created parallel to the neutral line (shear field). This suggests that flux emergence may be able to provide magnetic shear field sufficient to drive CMEs, but it is unclear how this magnetic shear emerges into the corona, as the axis of the emerging flux tubes remains rooted near the surface. 

 \citet{2004ApJ...610..588M} showed that shearing motions associated with the emergence are able to create a sheared coronal flux rope which rises out of the simulation domain, although the authors do not associate this with an eruption because the rise speed is limited by the amount of available 'axial flux' to drive an eruption.  Note that in this paper we use the term \textit{flux tube} to refer to the sub-surface tubes which we initiate our simulations with, and \textit{flux rope} to refer to the twisted structures formed in the corona. 3D simulations \citep{2009ApJ...697.1529F} have shown that without a pre-existing coronal field, the rise of coronal flux ropes created during flux emergence is constrained by the expanding emerging field of the original flux tube, or envelope field. In a similar manner, the work of \citet{2008ApJ...674L.113A}  and \citet{2009A&A...501..761M} showed that flux ropes formed from multiple flux tube emergence are ultimately trapped by the envelope fields. For an eruption to occur, the overlying tension of the envelope field must be removed.

\citet{2008A&A...492L..35A} added a pre-existing horizontal coronal field to their flux emergence simulations. They found that a coronal flux rope could be formed by inflows and horizontal shearing motions which drive reconnection within the expanding volume of the flux tube. They also found that reconnection between the envelope field and the pre-existing field reduced the overlying tension of the envelope field and allowed the flux rope to erupt. \citet{2009A&A...508..445M} found a similar result, but used a toroidal sub-surface flux tube which was already arched towards the surface, which allowed the flux tube axis to emerge bodily into the atmosphere.
\citet{2010A&A...514A..56A} showed that the likelihood of a coronal flux rope escaping when reconnection occurs between the ambient coronal field and the envelope field depends strongly on the strength of the ambient field.
 
 Flux emergence simulations such as these occur over timescales of 1000 s, much quicker than timescales of CME models, and do not extend high 
enough in the corona to allow a full comparison with CME models and observations.
The development of coupled models which span the timescales and spatial scales of both flux emergence and CME eruption is very important to fully understand the initiation of these explosive events.

In this work we investigate whether magnetic flux emergence, in a 2.5D cartesian geometry, can provide enough magnetic shear energy 
to drive CME initiation, with focus on the initiation mechanism of the magnetic breakout model. Figure \ref{fig:BO_energy_forces}(c) shows the energy in the shear magnetic field (normalized to the energy in the total field)   in a typical breakout model simulation as shearing is applied at the surface. Also shown is the total kinetic energy in the system. As the inner arcade is sheared at the $\beta=0.2$ lower boundary the shear energy increases. When the magnetic shear energy is about 20\% of the total magnetic energy, the upward magnetic pressure due to the shear field can force reconnection at the null point and hence further expansion, resulting in a rapid increase in kinetic energy. Hence the breakout model typically needs 20\% of the magnetic energy above the $\beta=0.2$ corona  to be in the shear field.
Although we know that flux emergence can create the shear flows and sheared magnetic
fields near the photosphere which are similar to those required in the low corona by CME models, it has not yet been calculated how much shear energy the emergence of magnetic flux can actually supply to the 
low $\beta$ corona. This issue is key to understanding how to couple dynamical flux emergence simulations and current CME simulations. We will investigate this using simulations of the evolution of sub-surface flux tubes with a variety of tube profiles as they emerge into both field-free corona and corona with pre-existing fields. We will also investigate whether flux emergence can drive CME initiation in these configurations.  

This paper is structured as follows: \S 2 explains the model, including numerical method, atmospheric and magnetic models, and initialization. 
\S 3 presents results
from 2.5D simulations. Firstly we 
look at emergence into a field-free corona, then we look at emergence into a quadrupole, and finally emergence into a dipole. In all cases we attempt to cover a realistic range of field profiles and coronal field strengths to test the robustness of our results.  \S 4 summarizes our findings and suggests consequences for modeling the initiation of CMEs by flux emergence.

\section{MODEL DESCRIPTION}

\subsection{Numerical Method}
We model our system of a magnetized plasma with a polytropic equation of state (polytropic index 
$\gamma=5/3$) using the resistive MHD equations. These are solved numerically using the Lagrangian 
remap code \textit{LareXd} \citep{2001JCoPh.171..151A}. The equations solved are presented below 
in Lagrangian form, using Gaussian units:
\begin{eqnarray}
\frac{D\rho}{Dt} & = & -\rho\nabla.\mathbf{v} \\
\frac{D\mathbf{v}}{Dt} & = & -\frac{1}{\rho}\nabla P 
+ \frac{1}{\rho}\frac{\mathbf{j}\wedge\mathbf{B}}{c} + \mathbf{g} + \frac{1}{\rho}\nabla.\mathbf{S}\\
\frac{D\mathbf{B}}{Dt} & = & (\mathbf{B}.\nabla)\mathbf{v} 
- \mathbf{B}(\nabla .\mathbf{v}) - c\nabla \wedge (\eta\mathbf{j}) \\
\frac{D\epsilon}{Dt} & = & -\frac{P}{\rho}\nabla .\mathbf{v}
+ \eta {j}^{2} + \varsigma_{ij}S_{ij}
-\frac{\epsilon-\epsilon_{0}(\rho)}{\tau}
\label{eqn:energy_MHD}
\end{eqnarray}
The gas density, pressure, and internal specific energy density  are denoted by 
$\rho$, $P$, and $\epsilon$ respectively and are 
defined at the center of each numerical cell.
The magnetic field, denoted by $\mathbf{B}$,  is defined at cell faces, $\mathbf{j}=c\nabla\wedge\mathbf{B}/4\pi$ is the current density, and $c$ is the speed of light. The
 velocity, $\mathbf{v}$, is defined at cell vertices. This staggered grid preserves 
$\nabla.\mathbf{B}$ during the simulation. The gravitational acceleration is denoted by $\mathbf{g}$, $\nu$ is the viscosity, set to $2.6\times{10}^{4} \ \textrm{g}.\textrm{cm}^{-1}\textrm{s}^{-1}$, $\eta$ is the resistivity, and 
 $\mathbf{S}$ is the stress tenor which has components 
$S_{ij}=\nu(\varsigma_{ij}-\frac{1}{3}\delta_{ij}\nabla.\mathbf{v})$, with
$\varsigma_{ij}=\frac{1}{2}(\frac{\partial v_{i}}{\partial x_{j}}+
\frac{\partial v_{j}}{\partial x_{i}}).$ The four equation are closed with a simple equation of state 
representing an ideal gas: $P=\rho R T$, where $R$ is the gas constant.

Equation \ref{eqn:energy_MHD}, minus the last term, $(\epsilon-\epsilon_{0}(\rho))/\tau$, is similar to that used in the majority of flux emergence simulations to date \citep{2008JGRA..11303S04A}. When flux tubes emerge and expand into the corona, the associated cooling due to the pressure terms can lead to unrealistic temperatures in the corona, as shown by \citet{2006A&A...450..805L}. \citet{2006A&A...450..805L} also showed that by including a relaxation term, such as the last term in Equation \ref{eqn:energy_MHD}, this unrealistic cooling could be avoided. This approach is a simple way to model the effects of terms that we are currently unable to formulate numerical equations for (such as chromospheric/coronal heating), and terms such as thermal conduction and radiative losses which are beyond the scope of this particular numerical code. The term $(\epsilon-\epsilon_{0}(\rho))/\tau$ relaxes the specific internal energy density back to its initial equilibrium values $\epsilon_{0}(\rho)$ on a timescale $\tau$ which scales with density:  
\be
\tau(y) = 23 {\left(\frac{\rho(y)}{\rho(y=0)}\right)}^{-1.43} \ \textrm{s}.
\ee

The equations are solved in 2.5D: the simulation box is 2D, with $x$ and $y$ being 
independent variables and $z$ being  ignorable, but all three components of the vector variables are evolved.
The simulation box extends from -3 Mm to 90 Mm in the vertical direction ($y$) and
from -45 Mm to 45 Mm in the horizontal direction ($x$). This is much larger than most flux emergence
simulations typically cover \citep{2004ApJ...610..588M,2009ApJ...697.1529F}, but is designed to allow expansion of the flux tubes into coronal field as far as possible. The numerical grid is stretched in both $x$ and $y$ to provide better resolution in areas of interest. The grid spacing, $\delta$, is smallest at the surface, where $\delta=0.0375 \ \textrm{Mm}$, 0.25 times the  scale height at the surface, and largest at the upper boundary, where $\delta = 0.15 \ \textrm{Mm}$.

In the majority of previous flux emergence simulations either ideal MHD is used ($\eta=0$), or $\eta$ 
is chosen to be a constant such that 
the diffusion due to this explicit resistivity is greater than the numerical diffusion in the code.
Theoretical evidence suggests that when the electron fluid flow speed, $v_{e}$, exceeds the phase speed
of the ion-acoustic model, $c_{ia}$, then ion-acoustic turbulence has a strong effect on current sheet development 
\citep{1988PhR...164..119B}.
In this case an effective formula for anomalous resistivity would be
\be
\eta =  \eta_{0}\max\left(0,\frac{v_{e}}{c_{ia}}-1\right)
\ee
which we can rewrite, using $v_{e} \sim |\mathbf{j}|/ne$, as
\be
\eta =  \eta_{0}\max\left(0,\frac{| \mathbf{j}|}{j_{crit}}-1\right)
\ee
where $j_{crit}=nec_{ia}$. We choose $\eta_{0} = 1.6 \times {10}^{-11}  \ \textrm{s} $, and $j_{crit} = 2.07\times{10}^{7} \ \textrm{statampere}.\textrm{cm}^{-2}$, and this anomalous  diffusion 
exceeds the numerical diffusion in the code. This approach has been used 
in MHD simulations of kink instabilities of coronal loops \citep{1999ApJ...517..990A,2001A&A...373.1089G}.

\subsection{Initial Conditions}

\subsubsection{Background Stratification} 
\begin{figure}
\begin{center}
\includegraphics[width=\textwidth]{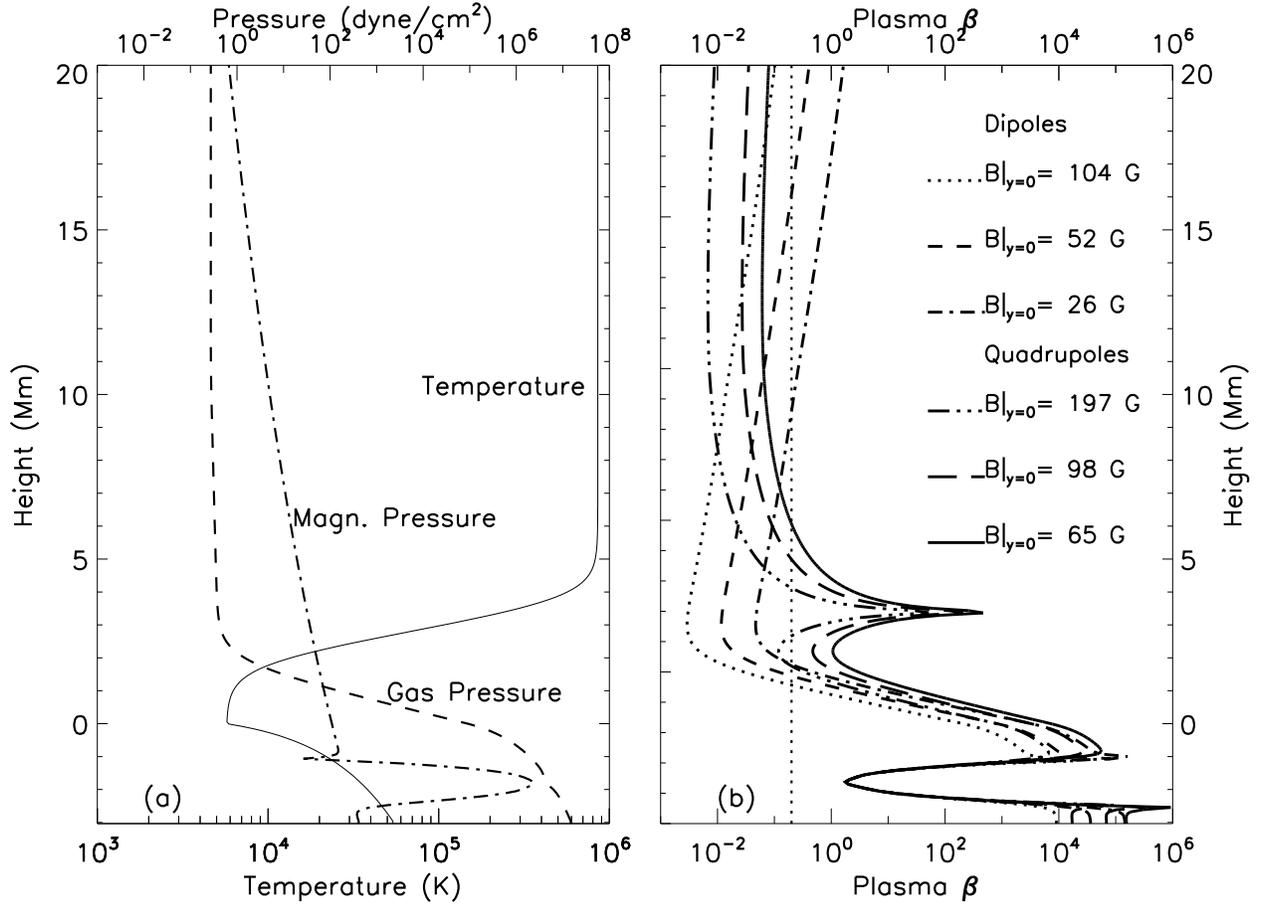}
\caption{Panel (a): the initial height profile of the model stratified atmosphere showing the temperature (solid line), gas pressure (dashed line), and magnetic pressure (dot-dashed line), up to 20 Mm. The magnetic pressure profile is for a sub-surface magnetic flux tube and an overlying, anti-parallel dipole. Panel (b): The initial plasma $\beta$ for six different simulations: Three simulations with background dipole field and three simulations with background quadrupole field. The straight black dotted line shows a constant value of $\beta=0.2$.
\label{fig:init_atm_beta}}
\end{center}
\end{figure}

The initial stratification is a simple 1D model of the temperature profile  of the Sun
which includes the upper 3 Mm of the convection zone, a photosphere/chromosphere, 
transition zone, and corona. The temperature profile is given by 
\be
T(y) = \left\{ 
\begin{array}{l l}
T_{ph} - \frac{|\mathbf{g}|}{R(m+1)}y, & \quad y\leq 0 \\
T_{ph} + \frac{(T_{cor}-T_{ph})}{2}
\left[\tanh \left(\frac{y-y_{tr}}{w_{tr}}\right)+1\right], & \quad  y>0 \\
\end{array}  \right.
\ee
The convection zone profile ($y\leq0$) is a linear polytrope which is marginally unstable to convection, with $m=\frac{1}{\gamma-1}$ being the adiabatic index for a polytrope.
The temperature in the photosphere and 
chromosphere is assumed to be constant, $T_{ph} =  5700$ K,
as is the temperature in the corona, $T_{cor} =8.6\times 10^{5} \ \textrm{K}$.
The height of the transition region is $y_{tr}=3.75 \ \textrm{Mm}$, and its width is $w_{tr}=0.75 \ \textrm{Mm}$. The density and gas pressure are specified by initially assuming hydrostatic equilibrium. The
resulting stratification is shown in Figure \ref{fig:init_atm_beta}(a).

\subsubsection{Magnetic Flux Tube}
We insert a cylindrical magnetic flux tube into the model convection zone at a height of 
$y_{t} =-1.8 \ \textrm{Mm}$. The axial field for all the tubes is given by
\be
B_{z} = B_{0} {e}^{-r^{2}/a^{2}}
\ee
where $r=\sqrt{x^{2}+(y-y_{t})^{2}}$ is the radial distance 
from the center of the tube, $B_{0} = 7800 \  \textrm{G}$ is the axial field strength at the center ($r=0$),
and $a=0.3 \ \textrm{Mm}$ is the width of the tube.  This axial field will equate to the shear field required by the breakout model.
We then choose three different tubes by choosing three different profiles for the twist
magnetic field ($B_{\theta}$ in the cylindrical coordinate system, $B_{x}$ and $B_{y}$ in cartesian). A minimum amount of twist is required to play the role of surface tension and keep the flux tube coherent as it rises in the convection zone \citep{1998ApJ...492..804E}. We denote the twist by
$\Theta$, defined as
\be
\Theta = \frac{B_{\theta}}{rB_{z}}.
\ee
The three twist profiles we use are given by
\begin{eqnarray}
\textrm{Tube 1}: \Theta_{1}(r) & = & c \\
\textrm{Tube 2}: \Theta_{2}(r) & = & c\left(1-e^{-r^{2}/c_{1}^{2}}\right) \\
\textrm{Tube 3}: \Theta_{3}(r) & = & ce^{-r^{2}/c_{2}^{2}}
\end{eqnarray}
where $c=1/a$, $c_{1}=0.15 \ \textrm{Mm}$ and $c_{2}=1.5 \ \textrm{Mm}$.

Tube 1, the constant twist tube, is similar to tubes which have been studied 
extensively in both 2D and 3D \citep{2001ApJ...554L.111F,2004ApJ...610..588M,2007ApJ...666..541A,2006A&A...450..805L,2009ApJ...697.1529F}. 
This initial condition has been shown to 
create a buoyant flux rope that can not only emerge into the 
solar atmosphere, but interact with a simple, horizontal pre-existing coronal field and create isolated flux ropes 
in the corona, and is thus an ideal choice for our study.

Tube 2 (increasing twist) and Tube 3 (decreasing twist), are similar to two tubes investigated by
\citet{2008A&A...479..567M} and have twist 
profiles which increase and decrease 
with radius, respectively. In their simulations investigating the effect of non-constant twist, the
tops of these two flux tubes emerged higher into the atmosphere than tubes with other twist 
profiles and are therefore the most relevant to this study. 

\begin{figure}
\begin{center}
\includegraphics[width=\textwidth]{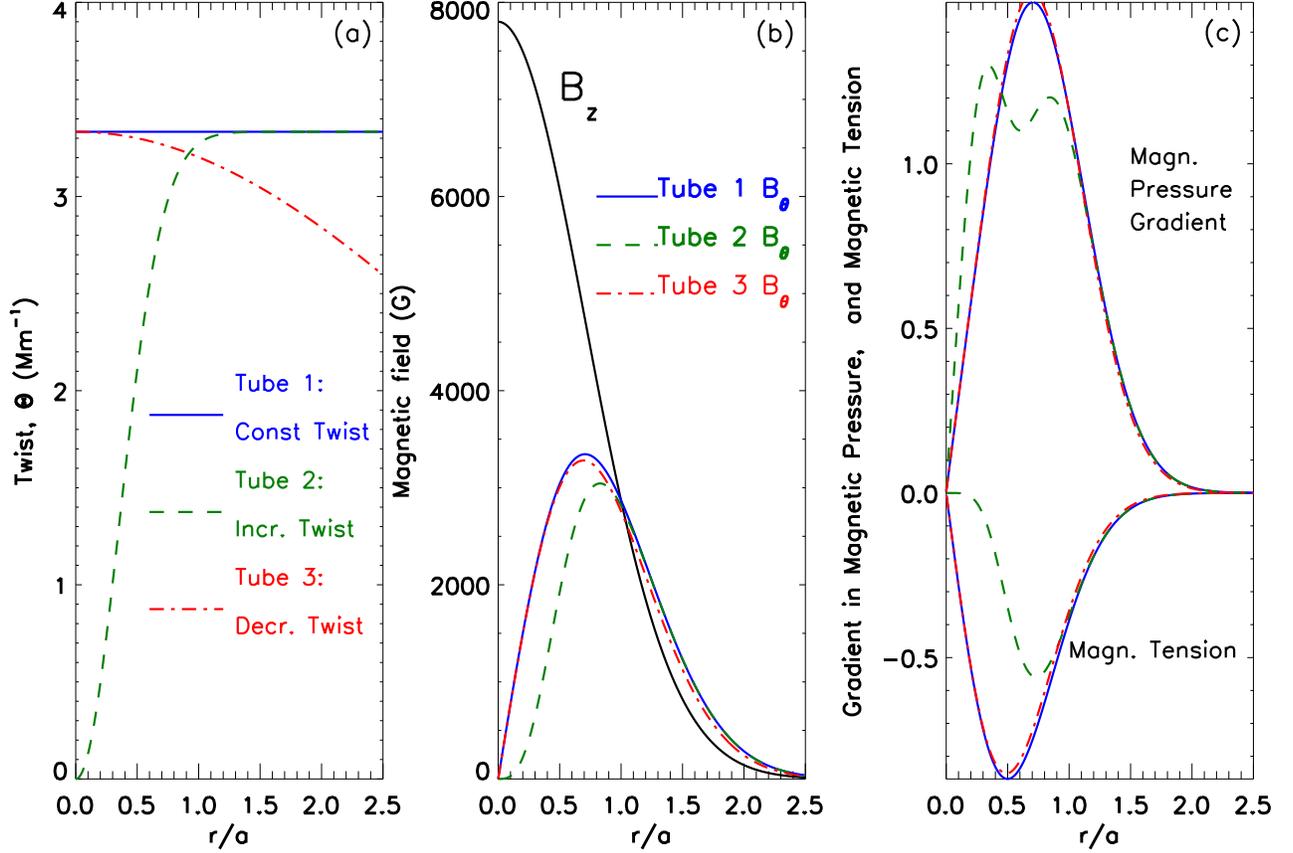}
\caption{Panel (a): Initial twist function, $\Theta$, for the three flux tubes. Tube 1: Constant twist, high tension, blue/solid lines. Tube 2: Increasing twist, low tension, green/dashed lines. Tube 3: Decreasing twist, high tension, red/dot-dashed lines. Panel (b): The axial field, $B_{z}$, and twist field, $B_{\theta}$, for the same three tubes. Panel (c):  The gradient in magnetic pressure,
$-\nabla (B^{2})/8\pi$ (dashed lines), and the magnetic tension, $B.\nabla B/4\pi$ (solid lines), both normalized by $B_{0}^{2}/a$, for the same three tubes. \label{fig:initial_twist}}
\end{center}
\end{figure}

As in \citet{2008A&A...479..567M}, we can group the three tubes into low magnetic tension and high magnetic tension cases.
The magnetic tension for these tubes is given by 
\be
\frac{\left(\mathbf{B}.\nabla \right)\mathbf{B}}{4\pi} = \frac{\left(-{B_{\theta}}^{2}/r\right) \hat{\mathbf{e}_{r}}}{4\pi}.
\ee 
The magnetic field, twist ($\Theta$), and tension profiles for the three tubes are shown in 
Figure \ref{fig:initial_twist}.
 Tubes 1 and 3 have similar twist and tension profiles. Tube 2 has lower tension near its axis due to the lower magnitude of $B_{\theta}$
close to the center of the tube. From their 3D simulations, \citet{2008A&A...479..567M} concluded 
that the twist profile was
not as important a factor in the emergence of the flux tubes as the axial field strength, which affected the rate of emergence but not the extent of the emergence. The purpose of choosing these three tubes in our study is to cover a range of flux tube behavior whilst 
keeping the potential amount of shear energy provided by flux emergence optimal. 

The plasma $\beta$ at the center of the flux tubes is approximately 10, and the flux tubes are all
non--force-free. To initiate a buoyant flux tube we perturb the background gas pressure by an 
amount $p_{1}$ such that $ \left(\nabla p_{1}\right)_{r} = \left(\mathbf{j}\wedge\mathbf{B}\right)_{r}$, which can be integrated to give
\be
p_{1}(r) = -\frac{B^{2}(r)}{8\pi} + \int_{r}^{\infty}\frac{B_{\theta}^{2}}{(4\pi r)}dr
\ee
so that the tube is in radial force balance. Assuming that 
the flux tube is in thermal equilibrium with its surroundings  makes the tube less dense that the surrounding plasma and initiates its buoyant rise to the surface. 

We will investigate the emergence of these tubes first into a field-free corona and then into different quadrupole and dipole coronal field configurations to test how efficient flux emergence is at providing magnetic shear energy to the corona.

\subsubsection{Quadrupole Background Magnetic Field}
\label{section:quadrupole}
\begin{figure}
\begin{center}
\includegraphics[width=\textwidth]{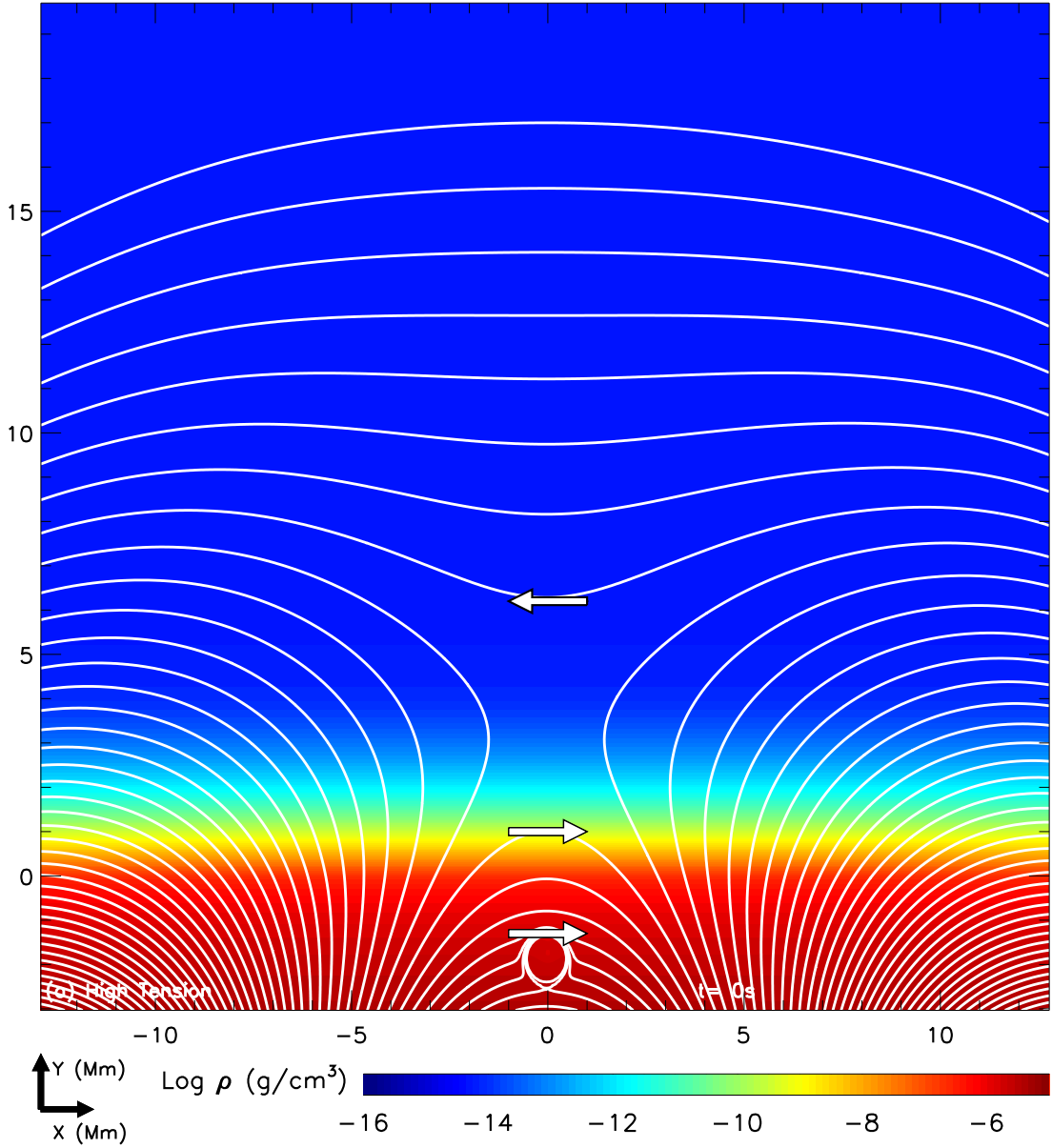}
\caption{Log of density at $t=0$ and initial magnetic field lines in the $x-y$ plane (constant contour intervals of the z component of the magnetic vector potential, $A_{z}$). The field lines show the outer part of the sub-surface flux tube (located at x=0, y=-2 Mm), and an overlying quadrupole field. The inner arcade of the quadrupole is parallel to the upper half of the flux tube, as can be seen by the arrows, which show the direction of $B_{x}$ in the plane. The null is located at about 3.2 Mm above the surface (y=0).
\label{fig:init_2D}}
\end{center}
\end{figure}

After investigating the emergence of flux tubes into a field free corona, we investigate the interaction of coronal field and emerging flux within the context of magnetic breakout. To do this we first impose a quadrupole coronal field above the magnetic flux tube.  We construct the quadrupole as in 
\citet{1996ApJ...460L..73K} by first defining the magnetic vector potential along a source surface at $y=-7.5Mm$ (which is outside the computational domain):
\be
A_{z}(x,-7.5 \ \textrm{Mm}) = q_{d}\left\{
\begin{array}{l l}
x^{2}\left[ 1-{(x/x_{a})}^{2} \right], & \quad 0 \leq |x| \leq x_{a} \\
0, & \quad  |x|>x_{a}  \\
\end{array}  \right.
\label{eq:az}
\ee
with $x_{a}=21 \ \textrm{Mm}$  being the horizontal extent of vertical qaudrupole field at this height. The vector potential in the interior of the domain is then determined by integration of Laplace's equation with Equation \ref{eq:az} as the lower boundary condition and $A_{z}=0$ at $y=\infty$. 
The quadrupole consists of an inner dipole and an overlying dipole of opposite orientation, separated by a null point at $y\sim3$ Mm.  The height of the null is chosen to allow for the expansion of the flux tube as it emerges into the lower corona, during which the cross-section of the tube increases by a factor of 10
 \citep{2007ApJ...666..541A}. Thus the 0.3 Mm flux tube will stay within the central arcade of the quadrupole when it emerges. The magnetic configuration of the sub-surface flux tube and overlying quadrupole  field is shown in Figure \ref{fig:init_2D}. For the breakout mechanism, most of the horizontal quadrupole flux above the null is removed by reconnection with horizontal flux below the null. 
 We therefore vary the strength of the quadrupole so that the horizontal flux above the null point, along the $x=0$ line, is a certain fraction of the horizontal flux contained between the center of the sub-surface flux tube and the null point. We choose $q_{d}$ to be [11.7, 5.85, 3.9]$\times 10^{7} \ \textrm{G}\textrm{cm}$, which give quadrupole surface strengths at $x=x_{a}/2$ of [197, 98, 66] G respectively. The horizontal flux above the null point is then  [1, 0.5, 0.3] times the horizontal flux between the flux tube and the null point. This allows us to control the amount of reconnection at the null point, whilst still being able to fix the $\beta$ at certain heights above the surface.  The curves in Figure \ref{fig:init_atm_beta}(b) shows the $\beta$ profiles for the configurations of a sub-surface flux tube plus three different strength quadrupoles. These profiles are generally consistent with $\beta$ profile models of active regions developed by \citet{1999SoPh..186..123G} and \citet{2001SoPh..203...71G} which take into account Soft X-ray Telescope limb observations. 
  
\subsubsection{Dipole BackGround Magnetic Field}
\label{section:dipole}
While emerging a flux tube into a quadrupole is the first step to coupling flux emergence and the breakout model, we also investigate a more self-consistent configuration, with a flux tube emerging into a background dipole field. By using a dipole with opposite orientation to the upper half of the flux tube, reconnection between these two systems creates a quadrupole structure self-consistently.

The dipole field is represented by the
vector potential $\mathbf{A} = A_{z}\mathbf{e}_{z}$ where
\be
A_{z}(x,y) = d\frac{y-y_{d}}{r_{1}^{3}},
\ee
with $r_{1}=\sqrt{x^2+(y-y_{d})^2}$ being the distance from the source. We chose $y_{d}$ to be -15 Mm so that the flux tube is far from the source of the 
dipole. To cover various dipole strengths we pick a range of d = [18, 9, 4.5]$\times 10^{28} \ \textrm{G}\textrm{cm}^3$
which gives a magnetic field strength at $x=0$,$y=0$ of [105, 53, 26] G respectively. 
In a similar fashion to the quadrupole field, we choose these
values so that the horizontal flux contained between the axis of the flux tube and the separatrix which separates the flux tube and the dipole is a certain fraction of the horizontal flux  above this separatrix. With decreasing dipole field strength, these factors are [1, 0.5, 0.25]. The $\beta$ profile for the three
dipole choices overlying a sub-surface flux tube are shown in Figure \ref{fig:init_atm_beta}(b). These choices of dipole strength allow for a range in the $\beta$ profiles, but are still consistent with the models of $\beta$ in the solar atmosphere developed by \citet{1999SoPh..186..123G} and \citet{ 2001SoPh..203...71G}.

\section{RESULTS}
\label{section:results}

\subsection{Initial Evolution in the Convection Zone and Lower Atmosphere: Effect of Twist Profiles}
\label{section:initial_rise}
The evolution of the flux tubes in the convection zone is similar to previous simulations of rising flux tubes 
\citep{1998ApJ...492..804E,2001ApJ...549..608M,2006A&A...450..805L}. The initial buoyancy causes the flux tube to rise to
 the surface, during which time the cross-section increases and flux conservation decreases the axial, or shear, field strength.
As the flux tube meets the 
convectively stable photosphere its rise is halted and horizontal expansion spreads the flux tube out to form a 
contact layer with the plasma above. The density above this layer is higher than the density below due to the concentration of magnetic field and the layer 
is unstable to a Rayleigh-Taylor like instability known as the magnetic buoyancy instability 
\citep{1961PhFl....4..391N,1961psc..book.....A,1979SoPh...62...23A}. The stability of the contact layer involves a competition between the destabilizing gradient in the magnetic field and the stabilizing sub-adiabatic temperature gradient
\citep{1970ApJ...162.1019G}.  The stabilizing term is dependent on the local $\beta$, and as the center of the tube rises up to this contact layer, the local $\beta$ falls and the
instability allows the upper portion of the flux tube to expand into the atmosphere, as in \citet{2004A&A...426.1047A}. 

\begin{figure}
\begin{center}
\includegraphics[width=\textwidth]{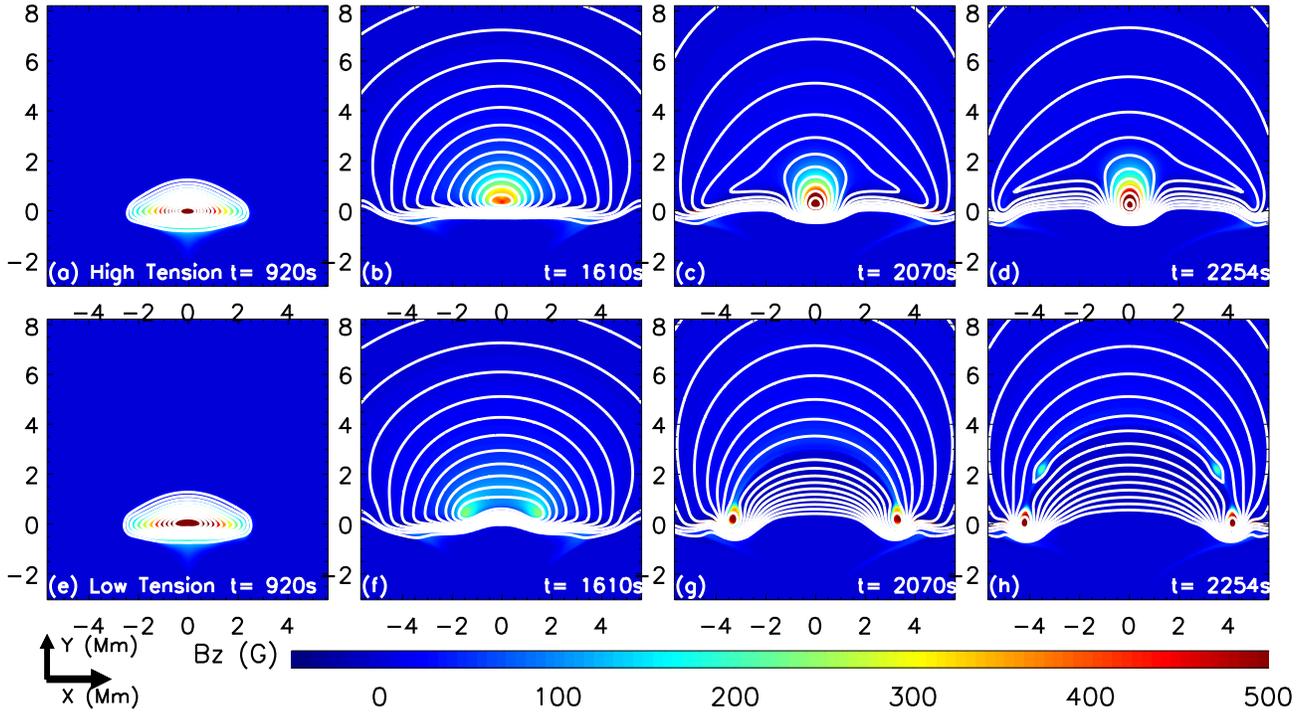} 
\caption{Panels (a) to (d): The magnetic shear field, Bz (G),  and magnetic field lines (constant contour intervals of $A_{z}$) at four different times  for the high tension flux tube (Tube 1) emerging into a field-free corona. Panels (e) to (h): The same but for the low tension tube (Tube 2). 
\label{fig:2a_4a_2D_0_98}}
\end{center}
\end{figure}

As can be seen in Figure \ref{fig:initial_twist}(c), the choice of twist profile affects the amount of tension in the tubes. The increasing twist tube (Tube 2) has relatively less tension near its axis than the constant twist tube (Tube 1) and the decreasing twist tube (Tube 3), and also a higher buoyancy at its center. This difference distinguishes the evolution of the increasing twist tube from the other two, and from now on we refer to the low tension case to mean Tube 2 and the high tension case to mean Tubes 1 and 3. In Figure \ref{fig:2a_4a_2D_0_98} we can see how the choice of twist profile affects the initial expansion into the field-free corona. The tubes act nearly identically as they rise buoyantly in the convection zone and also in the lower atmosphere, as the magnetic buoyancy instability develops and the outer fieldlines extend into the corona.  However, the two cases then differ in how the center of the tube reacts. In the high tension cases, the lower fieldlines near the center of the tube remain near the surface, shown in Figure \ref{fig:2a_4a_2D_0_98}(d), and the shear field is concentrated at the center, creating a single neutral line. In the low tension case, the decreased tension allows the center of the flux tube to rise, forming a crescent shape, shown in Figure \ref{fig:2a_4a_2D_0_98}(h). During this evolution, the mass in the center of the tube drains to the lowest possible location which is at either end of this crescent shape. The coupling of shear field and density in this 2.5D simulation means that the shear field also concentrates in these regions. Note that this emerging structure now has three neutral lines at the photosphere.

\subsection{Emergence into a Field-Free Corona: Effect of Twist Profiles}
\label{section:non_B_corona}

\begin{figure}
\begin{center}
\includegraphics[width=\textwidth]{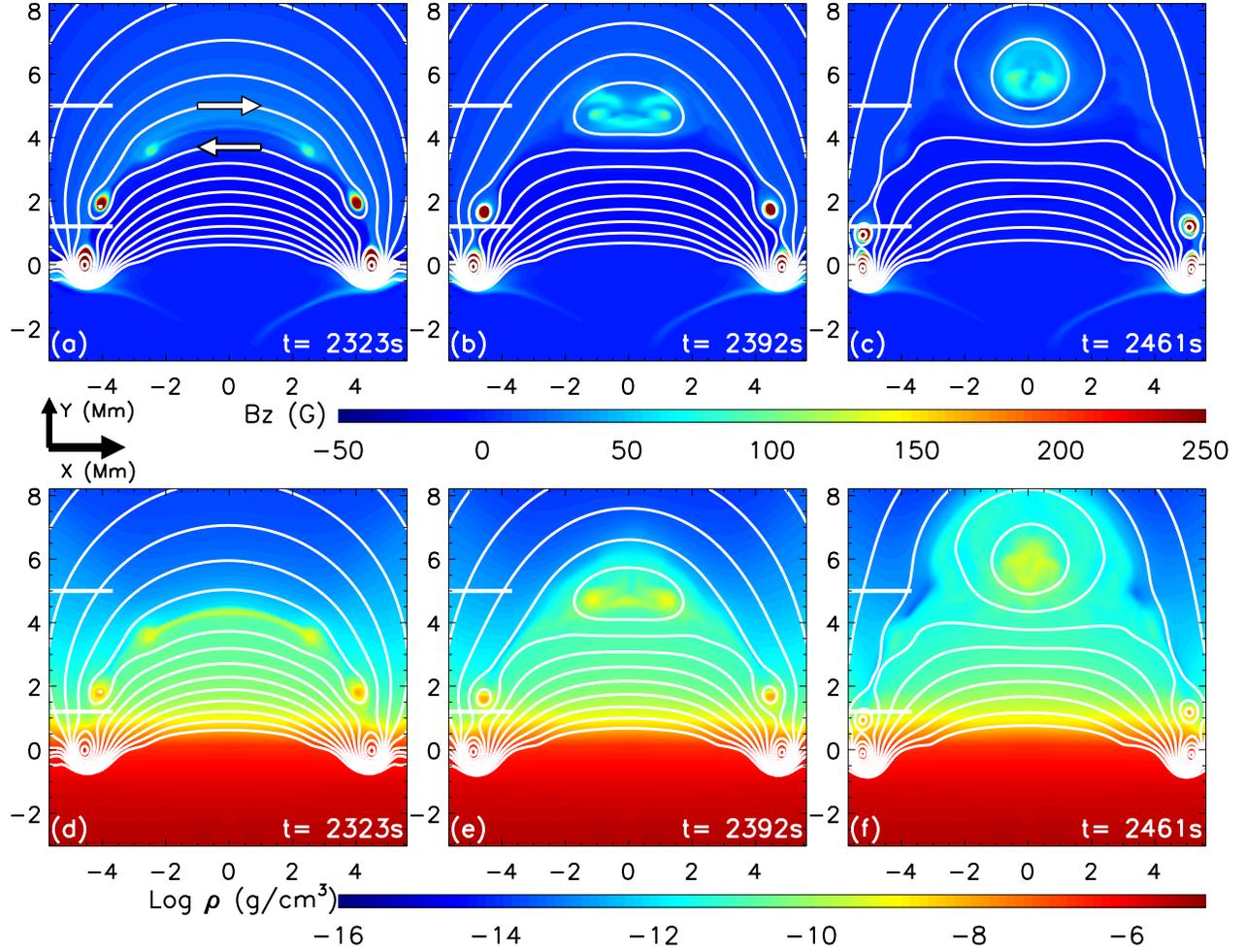}
\caption{Panels (a) to (c): The magnetic shear field, Bz (G),  and magnetic field lines, at three different times  for the low tension flux tube emerging into a field-free corona. Panels (d) to (f): The same but for the log of the density. The two white tick marks on the left are located at y=1.2 Mm and 5 Mm. \label{fig:4a_2D_Bsh_rho}}
\end{center}
\end{figure}

The center of the high tension flux tube remains at the surface for the duration of the simulation. 
We follow the continued evolution of the low tension flux tube in Figure \ref{fig:4a_2D_Bsh_rho}, which shows the shear field and the log of density, along with magnetic fieldlines.  The deformation of the center of the tube into a crescent shape forms a current sheet as field of opposite sign is brought together on either side of this crescent. This enhanced current triggers the anomalous resistivity and reconnection sets in. As can be seen from Figures  \ref{fig:4a_2D_Bsh_rho}(b) and (e), the reconnection sites are at either side of the apex of the current sheet arch. Plasmoids are formed which are expelled along the current sheet. The half width of the current sheet, $l$,  is approximately 0.15 Mm, and a typical distance between the plasmoids is 5 Mm. Thus the ratio of half current sheet width to the length between islands, $\lambda$, is approximately $l/\lambda=0.03$. Based on \cite{Priest2000}, the wavelength range for the tearing mode is 
\begin{equation}
\frac{1}{2\pi}{\left(\frac{D}{lv_{A}}\right)}^{1/4} < \frac{l}{\lambda} < 1/(2\pi) \sim 0.16
\end{equation}
where $D$ is the magnetic diffusivity ($D=\eta c^{2}/4\pi$) and $v_{A}$ is the Alfv\'{e}n speed. Using values from our simulations gives the lower limit as 0.01. Hence the current sheet dimensions, $l/\lambda=0.03$, are compatible with a tearing mode instability. Later in time, plasmoids are  formed closer to the apex of the current sheet and coalesce, creating an isolated flux rope at the center. Both shear field and mass are trapped in this flux rope, but the majority of the shear field concentrates at either end of the current sheet, in the low-lying dips in the magnetic field.

\begin{figure}
\begin{center}
\includegraphics[width=\textwidth]{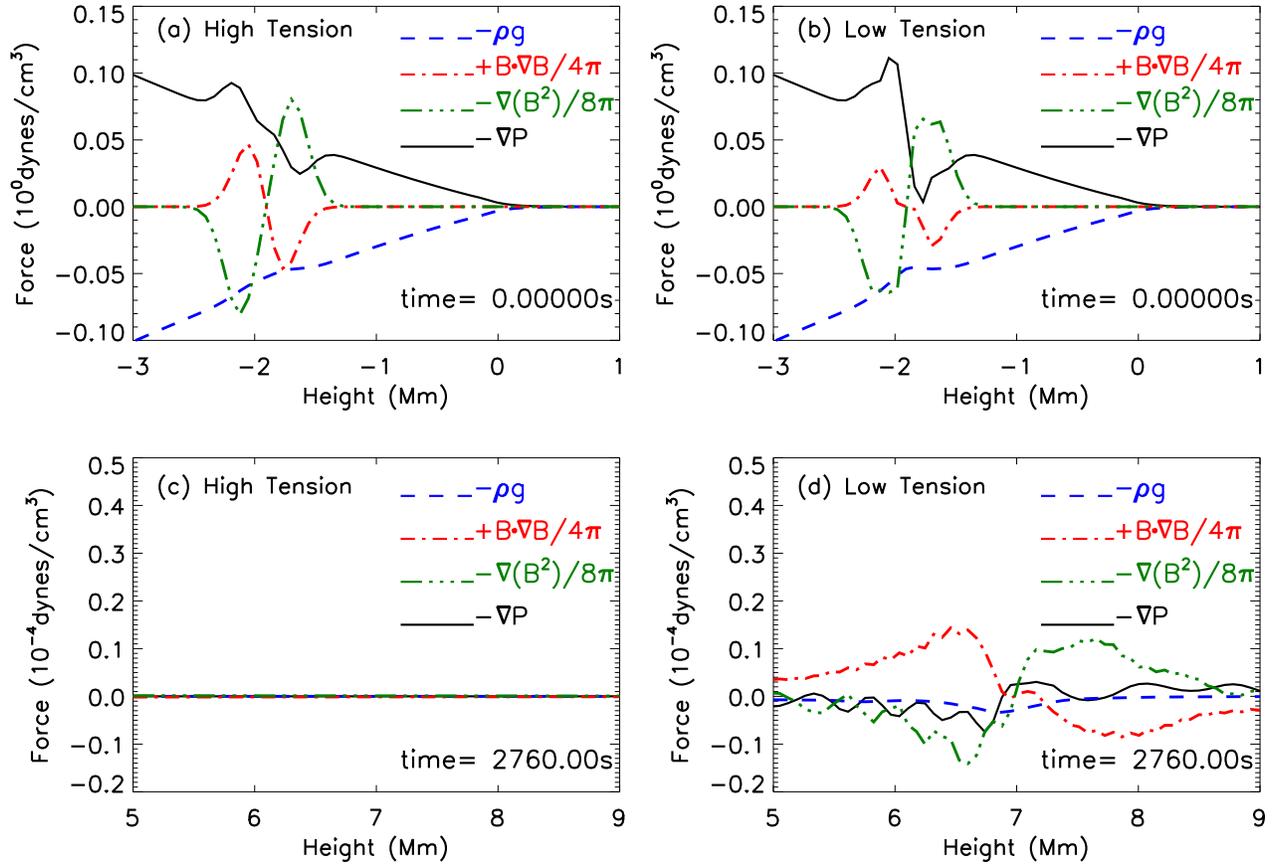} 
\caption{Panels (a) and (b): Vertical forces along the central, $x=0$, line inside the initial buoyant sub-surface flux tubes at  $t=0 \ \textrm{s}$ for the high tension and low tension cases. Panels (c) and (d): Vertical forces along the central, $x=0$, line above the surface at $t=2760$ s for the high tension and low tension cases. Panel (d) shows the coronal flux rope formed from the emerging low tension tube. 
\label{fig:2a_4a_forces}}
\end{center}
\end{figure}

To determine what happens to this coronal flux rope which is formed from the original center of the low tension flux tube, we can examine the forces in the flux rope. Figure \ref{fig:2a_4a_forces} shows the four vertical forces  associated with gravity, magnetic tension, magnetic pressure, and gas pressure along the central, $x=0$, line at two different times. The forces are shown for the initial sub-surface flux tubes in Figures \ref{fig:2a_4a_forces}(a) and (b), where the initial hydrostatic equilibrium can be seen in the large scale trends for $\rho\mathbf{g}$ and $-\nabla P$, and the magnetic forces can be seen over the range -2.5 Mm to -1.5 Mm. The outward directed magnetic pressure force, due primarily to the shear field, dominates over the
inward directed tension force of the twist field. The gas pressure force is directed inwards
to balance this excess magnetic pressure force. Figures \ref{fig:2a_4a_forces}(c) and (d) show the forces in the corona after the flux tube has emerged. For the low tension case, the forces in the newly formed coronal flux rope, the center of which is at $y=7 \ \textrm{Mm}$, can be seen in Figure \ref{fig:2a_4a_forces}(d).  The presence of mass can be detected in the small dip in the gravitational force.
Here, in contrast to the configuration of the initial flux tube, the
inward directed tension of the twist
field dominates over the outward directed magnetic pressure force. Thus
the gas pressure force is
directed {\it outward} here, to balance the excess tension force. This
shows that, in contrast to the breakout
model, shear field does not create a strong unbalanced upward/outward pressure force, and so there is no impetus which can drive a CME eruption. This shows the dramatic effect the dense lower solar atmosphere can have on the role of shear in driving CMEs.

As far as we are aware, the mechanism shown here for the creation of a coronal flux rope from an emerging flux tube whose cross section distorts into a crescent shape has not been seen before. We expect that such coronal flux ropes will form from 3D flux tubes which emerge with a large radius of curvature along their axis. A large radius of curvature implies that the draining of mass along the axis will not exceed the draining of mass perpendicular to the axis seen in these 2.5D simulations, and so the behavior will be similar. However, 3D simulations of similar flux tubes by  \citet{2008A&A...479..567M} do not show this behavior. We find that we still see the formation of our coronal flux ropes when we use the  same parameters ($y_{t}$=-0.17 Mm, $B_{0}$=3900 G, $a$=0.425 Mm, $c=3 \ \textrm{Mm}^{-1}$ and $c_{1}$=0.17 Mm) as one of  the low tension tubes used in \citet{2008A&A...479..567M}. 3D simulations are required to investigate this issue further.

\begin{figure}
\begin{center}
\includegraphics[width=\textwidth]{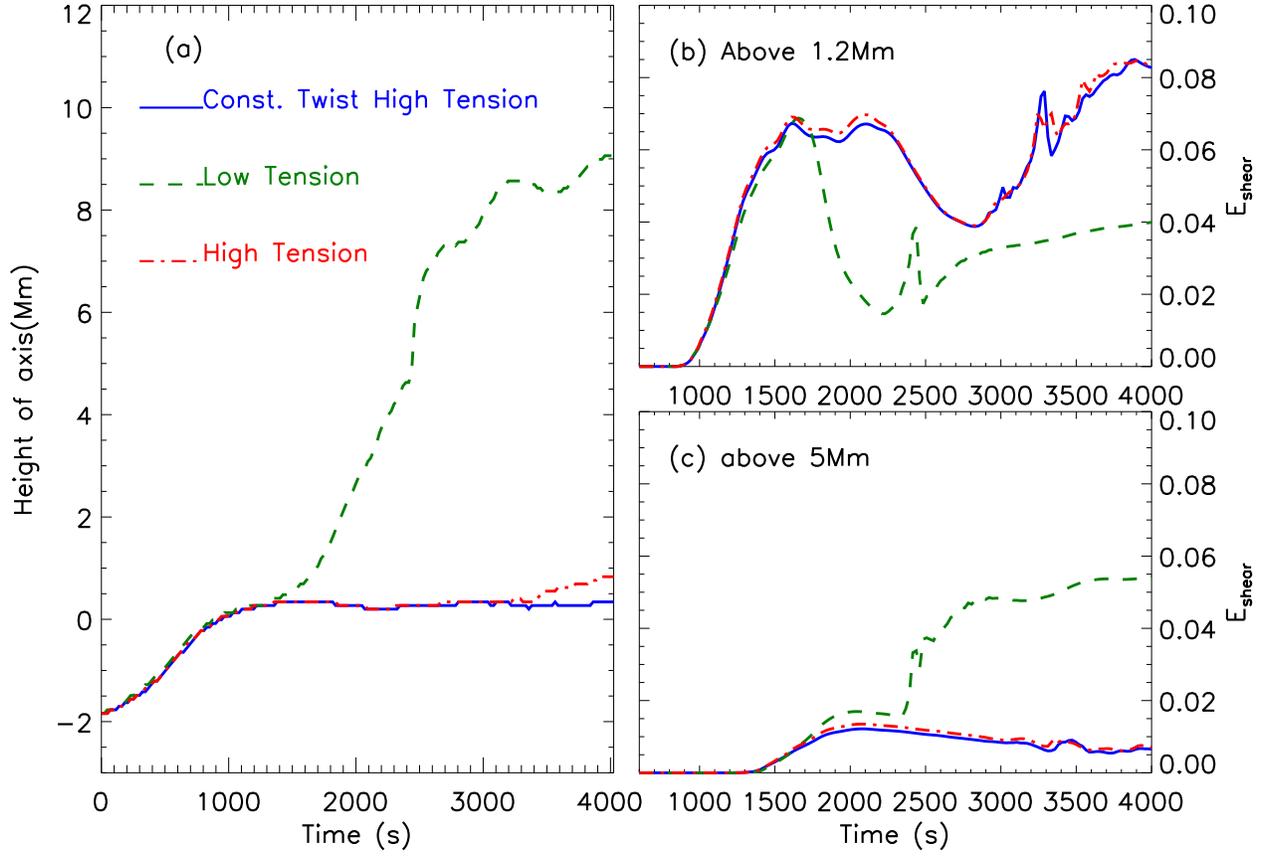} 
\caption{Panel (a): The height of the center of the three flux tubes emerging into a field-free corona.  Panels (b) and (c): The normalized magnetic shear energy, $\textit{E}_{\textrm{shear}}$ (see Equation \ref{eqn:shear}), above 1.2 Mm and 5 Mm respectively, for the same three tubes. 
\label{fig:2a_4a_5a_height_shear}}
\end{center}
\end{figure}

Figure \ref{fig:2a_4a_5a_height_shear}(a) shows the height of the centers of the three flux tubes as they emerge into the field-free corona. The center (which is initially the axis)  is defined at the point where $B_{x}|_{x=0}$ changes sign. The centers of the high tension tubes stay at the photosphere, while the center of the low tension tube rises to a height of about 8Mm as the tube is deformed into a crescent shaped current sheet and the resulting coronal flux rope rises. The coronal flux rope does not escape, most likely due to the presence of mass accumulation in the rope and due to the lack of magnetic shear. 

Figures  \ref{fig:2a_4a_5a_height_shear}(b) and (c) show the normalized magnetic shear energy above two different heights as a function of time for all three tubes. This normalized energy is given by
\be
E_{\textrm{shear}} = \frac{ \int_{y=y_{1}}^{y=90Mm}{\int_{x=-45Mm}^{x=45Mm}{B_{z}^{2}dxdy}    } }
                                             {\int_{y=y_{1}}^{y=90Mm}{\int_{x=-45Mm}^{x=45Mm}{|\mathbf{B}|^{2}dxdy} }    }        
\label{eqn:shear}
\ee
The first height,  $y_{1}=1.2 \ \textrm{Mm}$, is  the lowest height at which $\beta=0.2$ in the simulations in this paper with initial non-zero coronal field, and is an estimate of the height at which we need to transfer a significant amount of the shear magnetic energy
relative to the total magnetic energy in order to drive a CME. The second height is at $y_{1}=5 \ \textrm{Mm}$ above the surface and is at a location just below where the isolated flux rope is formed from the center of the low tension tube.

As the low tension case develops ($t$=1700 s to $t$=2200 s), the crescent shaped current sheet is formed and mass and shear field drain to the ends, below 1.2 Mm. Thus the shear energy above 1.2 Mm drops as shown by the green/dashed curve in Figure \ref{fig:2a_4a_5a_height_shear}(b). The coalescence of plasmoids, which leads to a concentration of shear flux, then creates an increase in $E_{shear}$ at $t$=2300 s. The expulsion of these plasmoids along the current sheet, down below 1.2 Mm, creates the sharp decrease seen shortly after at $t$=2400 s. As the flux rope formed from this current sheet then rises into the corona, carrying shear field with it, the shear energy above 5 Mm rises, as can be seen in the green/dashed plot in Figure \ref{fig:2a_4a_5a_height_shear}(c). The gradual rise in $E_{shear}$ for the high tension cases above 1.2 Mm after $t$=2800 s (red/dot-dashed and blue/solid plots in Figure  \ref{fig:2a_4a_5a_height_shear}(b)) is caused by reconnection at nearly vertical current sheets which form at the edges of the active region. This reconnection creates plasmoids which are able to transfer a small amount of shear above 1.2 Mm but which do not rise to the 5 Mm level.

The magnetic shear energy above 1.2 Mm for all three tubes does not exceed 10\% of the total magnetic energy above 1.2 Mm. For the emergence of a 2.5D flux tube into a field-free corona, the shear field is generally confined to the high $\beta$ lower atmosphere. The shear field which does rise in the coronal flux rope is not strong enough to create magnetic pressure which will counterbalance the magnetic tension and gravitational force in the rope and hence drive a further rise or an eruption. In the next section we look at the expansion into a quadrupole field, to see if the small amount of shear field which does emerge into the corona can drive reconnection at the null point of the quadrupole, thus reconfiguring the field to allow an eruption as in the breakout model.

\subsection{Emergence into a Quadrupole Coronal Field}
We present results of the emergence of the low tension flux tube into a pre-existing coronal quadrupole magnetic field. The quadrupole configuration can be destabilized by the presence of magnetic shear, as in the breakout model, but the effects of the lower atmosphere have not yet been investigated. We know from the previous section that the emergence of magnetic flux can bring shear into the corona, but for a field-free corona, the coronal flux ropes which then form are not sheared enough to erupt. In this section we emerge the low tension flux tube into a quadrupole of varying strength, and examine the interaction between emerging field and coronal field. We also look at the effect of the quadrupole strength on the overall emergence. We test whether the quadrupole configuration can be destabilized by flux emergence-supplied magnetic shear energy.  The initial configuration is shown in Figure  \ref{fig:init_2D} for the medium strength quadrupole ($B|_{x=x_{a}/2,y=0}=98$ G). 

\begin{figure}
\begin{center}
\includegraphics[width=\textwidth]{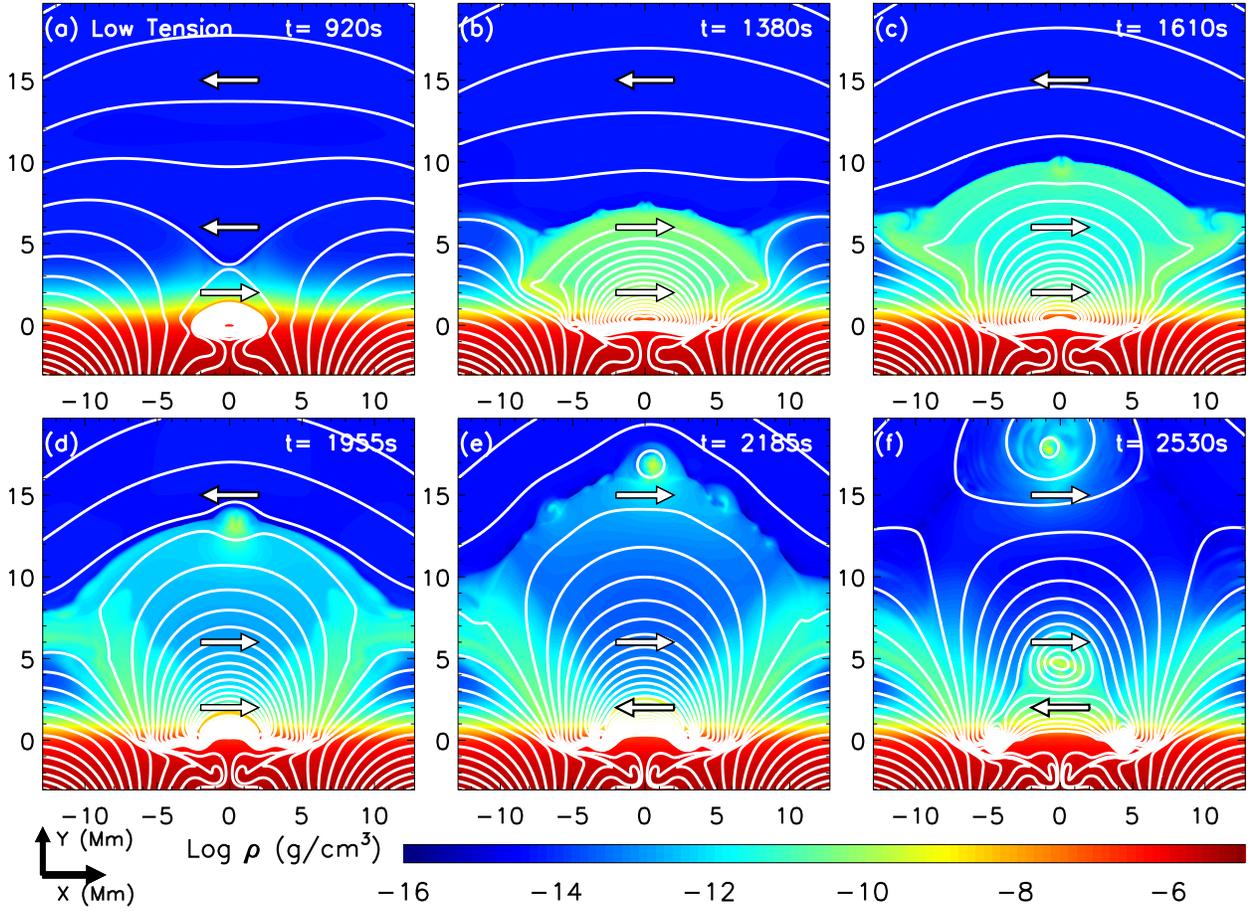} 
\caption{ Log of density and fieldlines for the low tension tube at six different times as it emerges into a medium strength ($B|_{x=x_{a}/2,y=0}=98$ G)  quadrupole field. The arrows show the direction of $B_{x}$ in the $x-y$ plane.
\label{fig:quad_2D}}
\end{center}
\end{figure}

The initial rise of the flux tube is similar to the field-free corona case up to about 1000 s. The presence of quadrupole field below the surface has a negligible effect on the rise of the tube, and its initial expansion into the atmosphere.  
Figure \ref{fig:quad_2D} shows the evolution of the low tension flux tube and the medium strength quadrupole at 6 different times (the coupling of $B_{z}$ and $\rho$ in these 2.5D simulations allows us to only show the density to demonstrate the structure of the coronal flux ropes). The field of the inner dipole of the quadrupole is aligned with the twist field of the upper part of the tube which expands into the corona. Thus the inner flux system of the quadrupole field is pushed upwards towards the null by the emergence, as shown in Figures  \ref{fig:quad_2D}(a) and (b). The expanding shell of the tube causes a crescent shaped current sheet to be formed as the original null is stretched over this shell. Reconnection sets in on either side of the apex of the current sheet. The current sheet continues to rise as the expanding shell of the emerging flux tube pushes it up as shown in Figures  \ref{fig:quad_2D}(d)-(f), while the reconnection creates a coronal flux rope at a height of 18 Mm.

In Figure \ref{fig:quad_2D} we can also see the deformation of the emerging flux tube's center, and the creation of a flux rope by reconnection at the current sheet formed at about 5 Mm, just as in the field-free corona simulations. Thus there are two coronal flux ropes formed, as can be seen in Figure \ref{fig:quad_2D}(f). The upper rope is created by the reconnection at the separatrix between the two systems of the quadrupole, driven by the expansion of the flux tube into the atmosphere. The lower rope is formed by reconnection at the current sheet formed from the center of the emerging flux tube. Both of these flux ropes carry shear field and mass.

\begin{figure}
\begin{center}
\includegraphics[width=\textwidth]{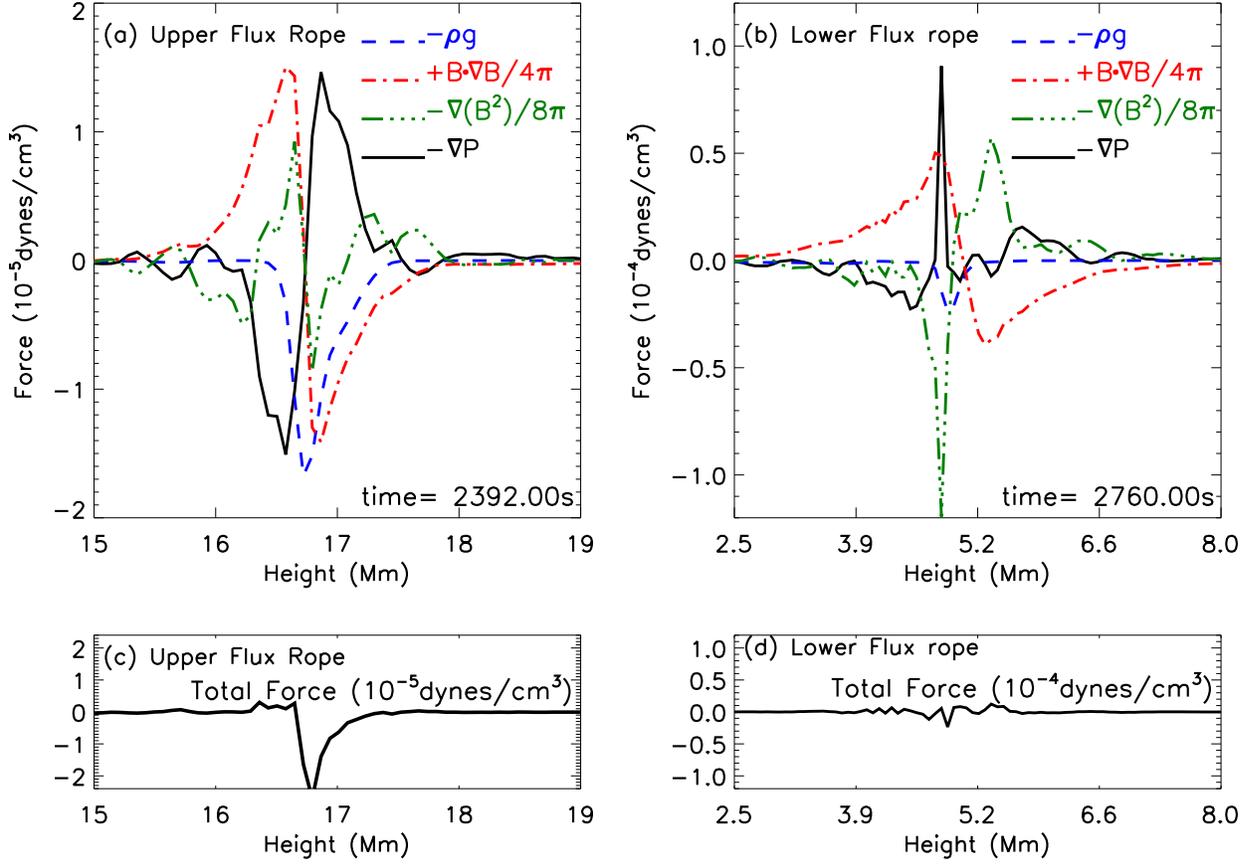} 
\caption{Vertical forces along the central, $x=0$, line for the two flux ropes formed when the low tension flux tube emerges into the medium strength  ($B|_{x=x_{a}/2,y=0}=98$ G) quadrupole. Panel (a) shows the forces in the upper flux rope formed by reconnection at the current sheet formed from the initial quadrupole null point. Panel (b) shows the forces for the flux rope created 
 by the deformation of the center of the emerging flux  tube. Panels (c) and (d) show the total vertical forces in these two flux ropes, respectvely. \label{fig:quad_forces_t_100.eps}}
\end{center}
\end{figure}

Figure \ref{fig:quad_forces_t_100.eps} shows the vertical forces in these two coronal flux ropes at later times of $t=2392 \ \textrm{s}$ and $t=2760 \ \textrm{s}$. Both ropes carry significant amounts of mass and  in both ropes  the magnetic tension acts inwards, counteracting any outward force due to gas or magnetic pressure. Figure \ref{fig:quad_forces_t_100.eps}(a) shows that {\it both} the magnetic tension and magnetic pressure forces are directed inwards in the upper flux rope, thus a strong outward pressure force is required for force balance. In
addition, the downward gravity force of the trapped mass is as strong as any of the pressure or
tension forces. These other
forces all have an upward directed asymmetry to help balance this large
gravity force, but nonetheless, as can be seen in Figure \ref{fig:quad_forces_t_100.eps}(c), they cannot and 
the flux rope starts to sink. In contrast, the gravity force is
relatively small in the lower flux rope,
as shown in \ref{fig:quad_forces_t_100.eps}(b), and the magnetic  pressure force is directed
outwards. Yet this outward
magnetic pressure force is well balanced by the inward magnetic tension.
Thus, while the ropes
differ in the way the magnetic and gas pressures act, the key fact is that the amount of shear field in them is  not enough to create an unbalanced outward acting magnetic pressure force to cause any further expansion. 

\begin{figure}
\begin{center}
\includegraphics[width=\textwidth]{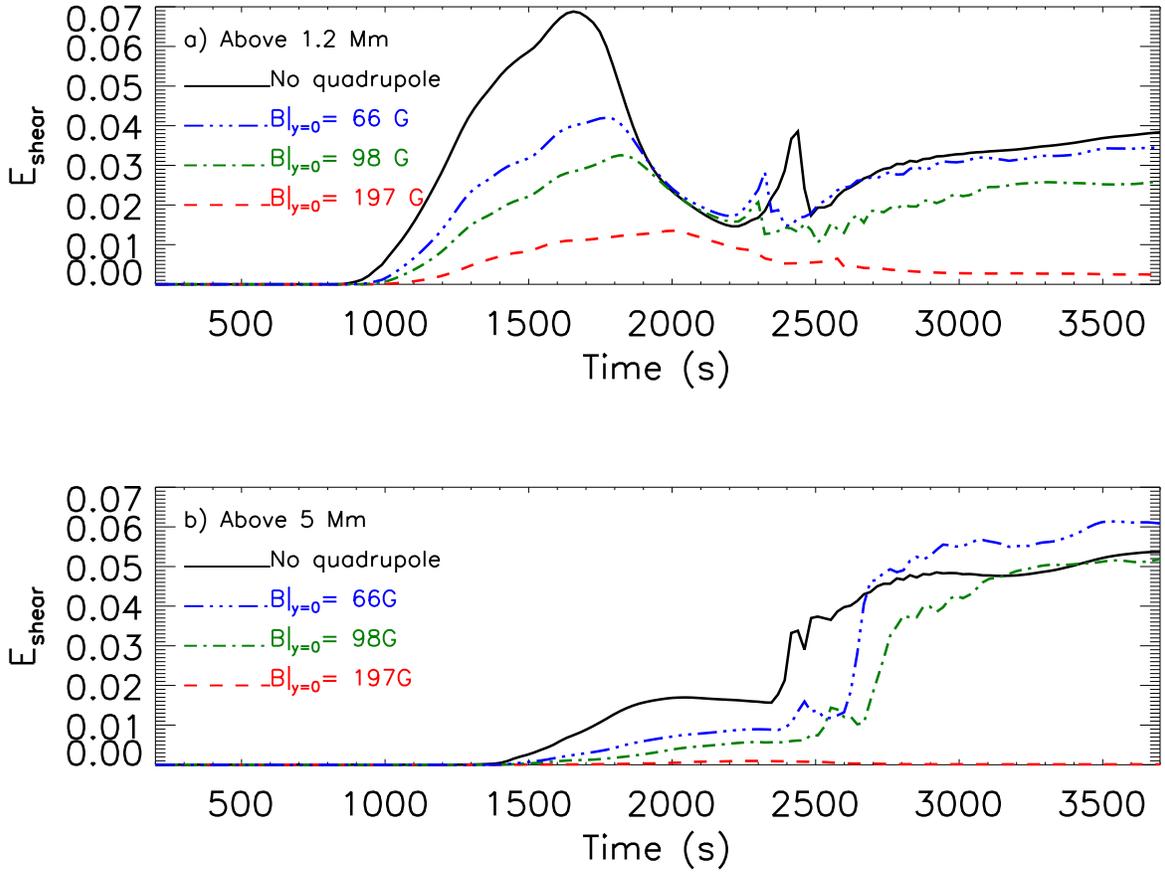} 
\caption{Normalized magnetic shear energy, $\textrm{E}_{\textrm{shear}}$ (see Equation \ref{eqn:shear}), above two different heights for the low tension flux tube emerging into a quadrupolar coronal field. Panel (a) is for 1.2 Mm. Panel (b) is for 5 Mm.
\label{fig:quad_shear}}
\end{center}
\end{figure}

The fraction of the magnetic energy contained in the shear field above the heights of 1.2 Mm and 5 Mm for the three choices of quadrupole and the field-free corona case are shown in Figure \ref{fig:quad_shear}. In general, increasing the strength of the quadrupole decreases this fraction as the total magnetic energy in the system increases. However this also increases the amount of reconnection at the separatrix of the quadrupole as the configuration contains  more flux above and below the null.

We have shown that the emergence of a flux tube into a quadrupole field does cause reconnection at the separatrix between the two flux systems in the quadrupole. However, the nature of the reconnection means that it does not remove significant amounts of horizontal field from above and below the original null point, as in the single X-point reconnection seen in the breakout model. The reconnection at the null is driven here by expansion of the outer part of the emerging flux tube, and the current sheet formed prefers reconnection off-center, creating a coronal flux rope which does not have enough magnetic pressure to escape.  This flux rope eventually falls back down, and joins with the flux rope which was created from the center of the emerging low tension flux tube. The magnetic energy in the shear field above the $\beta=0.2$ level is too small (less than 10\% of the total magnetic energy) to provide the shear energy required by the breakout model, which is consistent with the fact that the shear field in the flux ropes is not strong enough to overcome the tension and gravitational forces. 

 In the next section we show the emergence of flux tubes into a dipole coronal field which is orientated opposite to the upper part of the emerging flux tube. Reconnection between the expanding tube and the pre-existing dipole coronal field may remove twist field in the coronal flux ropes formed and allow them to rise further. 

\subsection{Emergence into a Dipole Coronal Field: Effect of Tension Profile}
\label{section:tubes_dip2}
 
In this section we present results from simulations of the emergence of the three different flux tubes into a particular dipole field to highlight the importance of the choice of 
tube profile. We choose the dipole field with $B|_{x=0,y=0}=52 \ \textrm{G}$. This choice of dipole best highlights
the differences between the high and low tension cases. We will investigate the choice of dipole strength in \S \ref{section:tube2_dips}.

\begin{figure}
\begin{center}
\includegraphics[width=\textwidth]{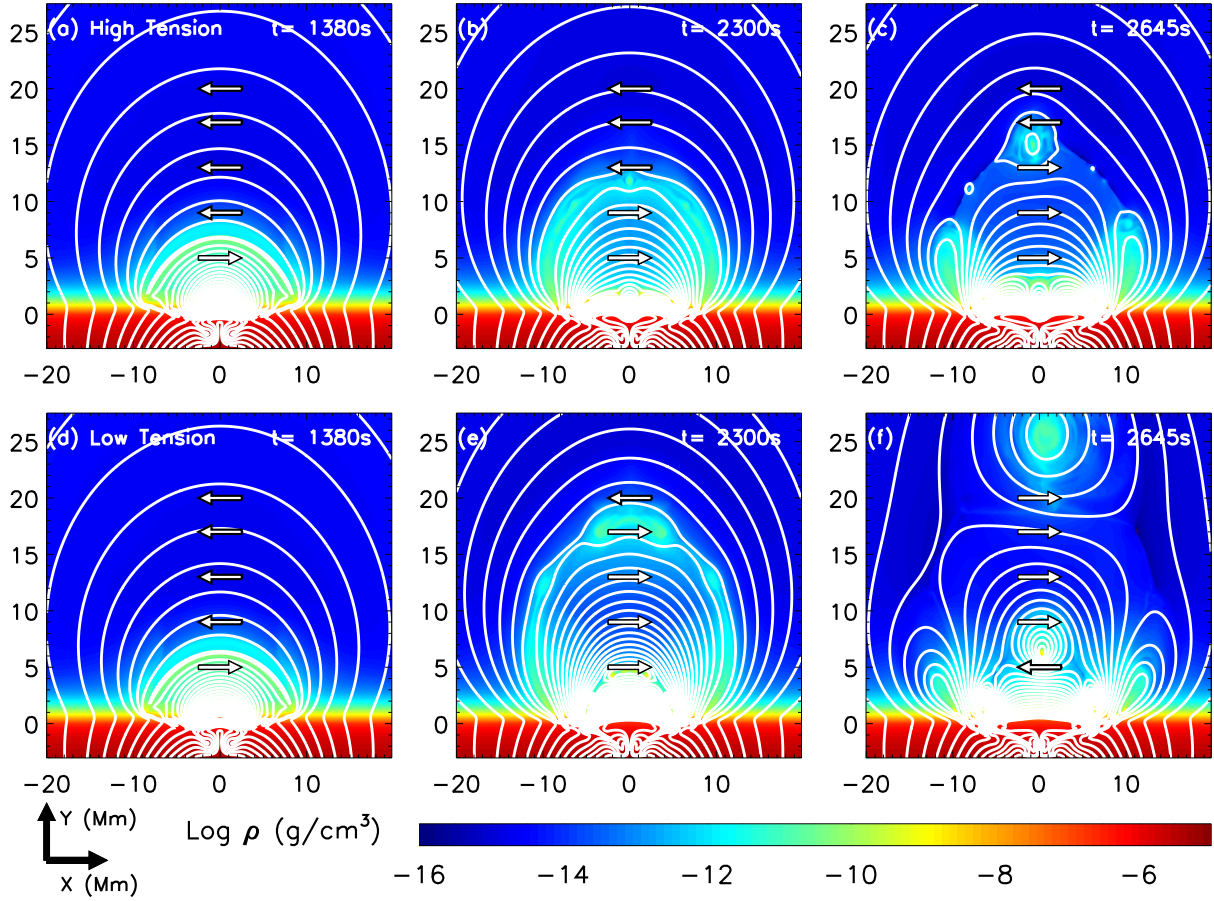} 
\caption{Log of density and fieldlines at three different times for the emergence of a high tension flux tube, panels (a) to (c), and a low tension flux tube, panels (d) to (f), as they emerge into the medium strength, $B|_{x=0,y=0}=52$ G, dipole coronal field. The arrows show the direction of the magnetic field.
\label{fig:2c_4c_flux_ropes}}
\end{center}
\end{figure}

Figure \ref{fig:2c_4c_flux_ropes} shows the emergence of the high and low tension tubes into the dipole field. The behavior of the flux tubes prior to emergence is very similar to the field-free corona and quadrupole coronal field cases. The outer part of the flux tube and the coronal dipole field lie on either side of a separatrix. As the flux tubes rise into the dipole field this separatrix is pushed upwards and becomes a current sheet, as shown in Figures \ref{fig:2c_4c_flux_ropes}(a) and (d). In both cases the current sheet undergoes reconnection, consistent with the tearing mode instability, creating plasmoids on either side of the apex which are expelled along the current sheet. Eventually a coronal flux rope is formed by coalescence of plasmoids near the apex of the current sheet. This flux rope has an initial rise speed of 25 km/s. In the low tension case, the flux rope is formed more rapidly and at a higher point in the corona. Also, a flux rope is formed in the lower corona by the deformation of the center of the emerging low tension flux tube, leaving two flux ropes in the corona. The structure of these two flux ropes is similar to that of the flux ropes formed in the field-free and quadrupole coronal field experiment, with magnetic tension and a gravitational force acting to prevent any further rise or expansion.  The creation of these coronal flux ropes appears to be the tearing mode instability, as was the case in simulations of flux emergence into a pre-existing field performed by \citet{2006ApJ...645L.161A, 2007A&A...466..367A}. However this is a different mechanism from that which creates the flux ropes in the simulations of \citet{2004ApJ...610..588M}, \citet{2008A&A...492L..35A}, and \citet{2009ApJ...697.1529F}, which is based on shear flows, converging motions, and internal reconnection of an emerging flux tube. 

\begin{figure}
\begin{center}
\includegraphics[width=\textwidth]{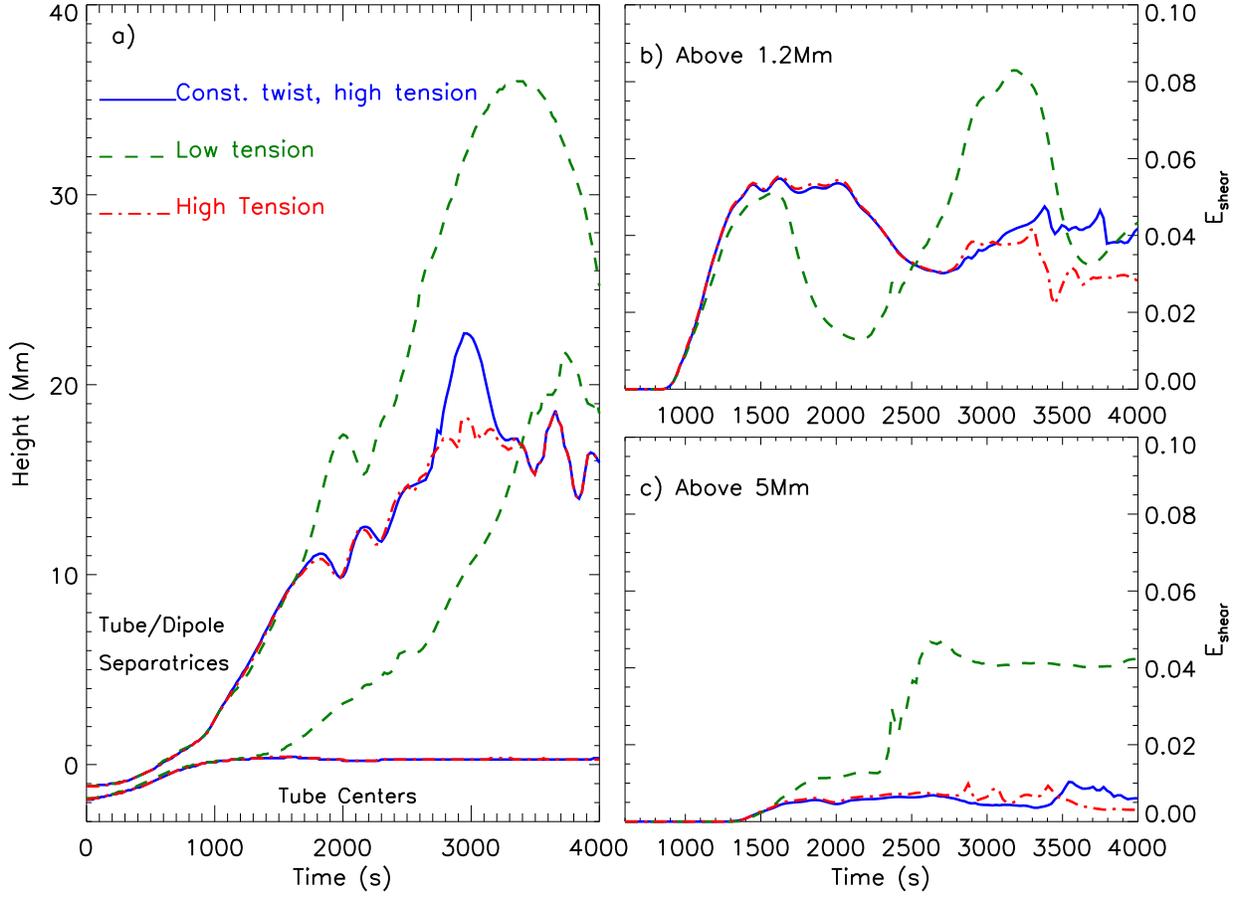} 
\caption{ Panel (a): Height of the flux tube centers (dotted lines) and tube/dipole separatrices (solid lines) for the three flux tubes emerging into a dipole of medium strength ($B|_{x=0,y=0}=52$ G).  Panels(b) and (c): The normalized magnetic shear energy, $\textit{E}_{\textit{shear}}$ (see Equation \ref{eqn:shear}), above heights 1.2 Mm and 5 Mm for the same three experiments.
\label{fig:dip_tubes_height_shear}}
\end{center}
\end{figure}

Figure \ref{fig:dip_tubes_height_shear}(a) shows the height of the flux tube centers (solid lines) and the  height of the separatrices between flux tube and dipole (dashed lines), along the $x=0$ line for the three tubes. Initially the paths of the centers and separatrices  are the same for all three tubes as they expand and push the dipole field upwards. After $t=1500$ s the low tension tube undergoes reconnection at its center, forming the lower of the two flux ropes.  Reconnection at the separatrix between the emerging tube and  the dipole field in all three simulations creates the coronal rope higher in the corona (solid lines). These coronal flux ropes eventually fall back down through the transition region after $t = 4000$ s. 

The amount of normalized magnetic shear energy supplied to the two heights of 1.2 Mm and 5 Mm for all three tubes emerging into the dipole can be seen in 
Figures \ref{fig:dip_tubes_height_shear}(b) and \ref{fig:dip_tubes_height_shear}(c).  The normalized shear energy above 1.2 Mm increases after 2000 s for the experiment where the low tension tube emerges into the dipole, shown as the green/dashed curves on Figure \ref{fig:dip_tubes_height_shear}(b). This is caused by the removal of some horizontal flux by reconnection in the current sheets. While this is similar to the reconnection which leads to an eruption in the breakout model, this reconnection does not remove enough horizontal field to raise the importance of the shear field in the coronal flux ropes formed by this reconnection. As a result the shear magnetic energy is still less than 10\% of the total magnetic energy above the $\beta=0.2$ corona. 3D simulations by \citet{2008A&A...492L..35A} showed that reconnection between emerging flux tubes and horizontal coronal field allows coronal flux ropes which are formed during flux emergence to escape. This is not the case in these 2.5D simulations with a dipole coronal field. In the next section we investigate the importance of the dipole strength on the behavior of the coronal flux ropes.

\subsection{Emergence into a Dipole Coronal FIeld: Effect of dipole field}
\label{section:tube2_dips}

By varying the dipole strength we can control the location and amount of reconnection at the current sheet formed between the flux tube and the dipole, and through this affect the amount of shear energy (relative to total magnetic energy) which is transferred to the  low $\beta$ corona. 
In this section we present results of the emergence of the low tension flux tube into a range of dipoles. These dipoles have horizontal fluxes which are 
factors of the horizontal flux that is initially in the top half of the emerging tube of 1, 0.5 and 0.25. We also compare
these three cases to the field-free corona case.

\begin{figure}
\begin{center}
\includegraphics[width=\textwidth]{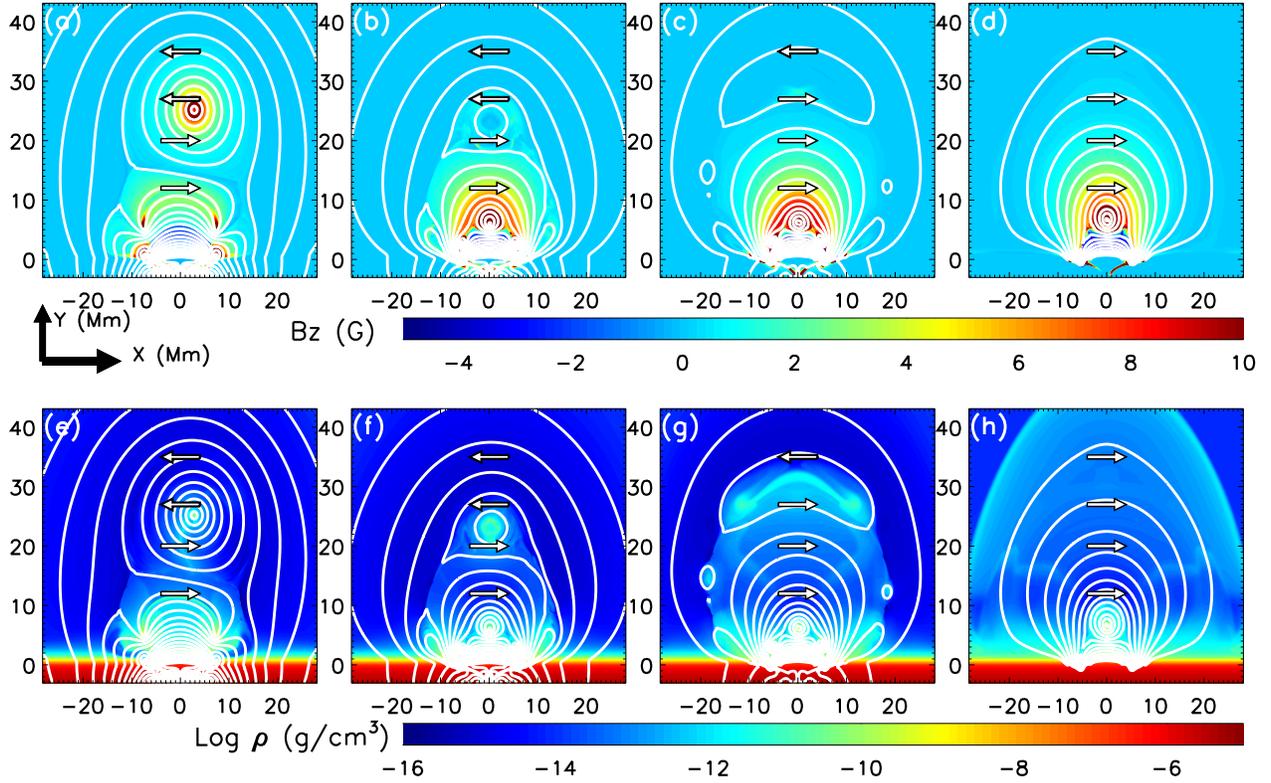} 
\caption{Shear field, $B_{z}$(G), and fieldlines  at t=2530 s for the emergence of the low tension flux tube into 4 coronal fields of decreasing strength. Panel a): $B|_{x=0,y=0}=104$ G. Panel b): $B|_{x=0,y=0}=52$ G. Panel c): $B|_{x=0,y=0}=26$ G. 
Panel d): $B|_{x=0,y=0}=0$ G (no dipole). Panels (e) to (h) show the same but for the log of density. Arrows show direction of magnetic field.
\label{fig:dips_2D}}
\end{center}
\end{figure}

The strength of the dipole affects the expansion of the outer part of the flux tube, as the stronger the dipole is the more tension it has to resist the expansion of the upper part of the flux tube into the corona.  The stronger the dipole, the quicker (and lower down in the corona) the two systems are forced together by the emergence and so the quicker the reconnection occurs between them and the quicker the coronal flux rope is formed. This can be seen in Figure \ref{fig:dips_2D} which shows snapshots at t=2530 s for the low tension flux tube emerging into the 3 different dipoles (of decreasing strength from left to right) and a field-free corona (far right panel). 
At this time, the stronger dipole case has already formed a coronal flux rope, while the weaker dipole case has only just started reconnection between the emerging flux tube and the dipole.

\begin{figure}
\begin{center}
\includegraphics[width=\textwidth]{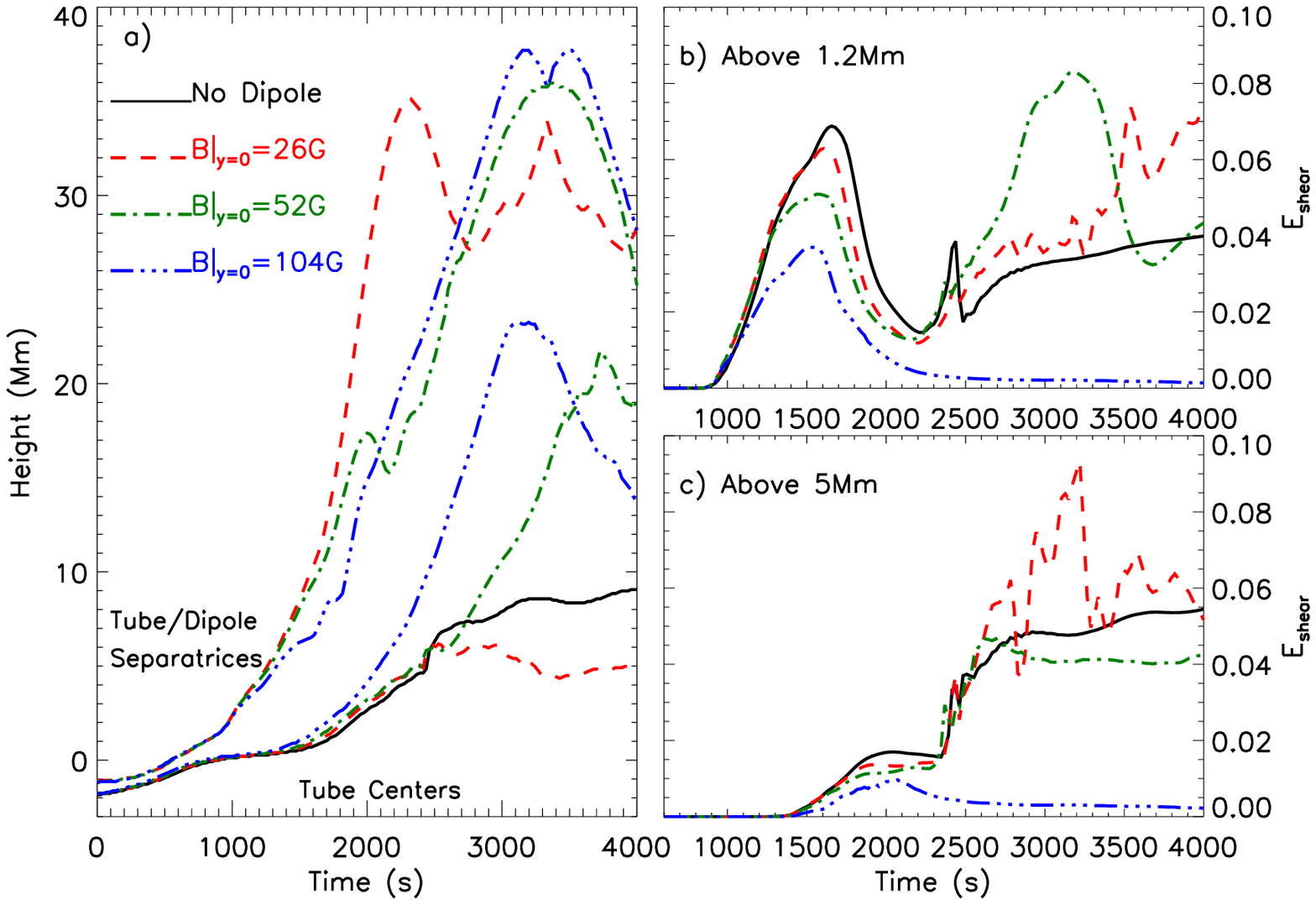} 
\caption{ Panel (a): Height of the flux tube centers, (dashed lines) and tube/dipole separatrices (solid lines) for the low tension flux tube emerging into dipoles of varying strength. Panels (b) and (c): The normalized magnetic shear energy, $\textit{E}_{\textrm{shear}}$ (see Equation \ref{eqn:shear}), above heights 1.2 Mm and 5 Mm for the same experiments.
 \label{fig:dips_heights_shear}}
\end{center}
\end{figure}

Figure \ref{fig:dips_heights_shear}(a) shows the heights of the flux tube centers and the flux tube/dipole separatrices  for the different choices in dipole strength. The separatrices all reach similar heights, but do so at different times, due to the differing tension in the dipoles which resists the pushing up of the dipole by the expanding flux tube. 

Figures \ref{fig:dips_heights_shear}(b) and \ref{fig:dips_heights_shear}(c) show the normalized shear energy above 1.2 Mm and 5 Mm. In general, the magnetic shear energy due to flux emergence is confined too low in the atmosphere to have a strong enough effect on the magnetic field in the corona. The shear energy is still always below 10\% of the total energy above the low $\beta$ corona. Any shear field that is transported to the corona by coronal flux ropes is too small to affect the twist field at these heights. The coronal flux ropes formed are not sheared enough and contain  significant amounts of dense material. In order to drive a CME by flux emergence in these simulations, we need a method  for transferring more shear field to coronal heights while transferring less mass. 

\section{CONCLUSIONS}

We have investigated the initiation of CMEs by dynamical flux emergence using 2.5D cartesian simulations of the emergence of a range of sub-surface flux tubes into a range of coronal magnetic field configurations. Most current CME models do not include the lower dense atmosphere in their models, but the boundary conditions for the shear energy and  dynamical flows they use are most likely results of flux emergence through this region. In a first attempt to couple the modeling of CME initiation and flux emergence, we have taken coronal field configurations which are based on the magnetic breakout model. However, 
our results are generally applicable to any CME model which assumes the appearance of highly sheared flux at the low $\beta$ corona.

The evolution of flux tubes into a field-free corona can be separated into low and high tension cases, as in \citet{2008A&A...479..567M},  based on the relative amount of tension near the center of the flux tube. During the emergence of a low tension tube,  an isolated flux rope was created by the deformation of the flux tube center  into 
a current sheet and the reconnection of fieldlines on either side of the apex of this current sheet.  Since twisted flux ropes are essential for several CME models \citep{1999A&A...351..707T,2003ApJ...588L..45R,2006PhRvL..96y5002K,2007ApJ...668.1232F} further investigations are warranted, especially in 3D, to take into account the effect of curvature along the tube's axis.   The comparison of 3D simulations with observations of neutral lines during flux emergence will be able to verify if this behavior is occurring on the Sun.

In an attempt to test whether the shear field brought into the corona by flux emergence could destabilize the quadrupole configuration of the magnetic breakout model, we added a pre-existing quadrupole coronal field above the emerging flux tubes. The expanding outer shell of the flux tube stretched the original null point of the quadrupole into a current sheet, and reconnection in this current sheet created another coronal flux rope, higher in the corona. Both types of flux ropes were created in similar fashions: by the formation of a crescent shaped current sheet, with preferential reconnection occurring off-center, forming a coronal flux rope at the apex
 of the crescent. However, one type of flux rope was created by the deformation of the center of an emerging flux tube, and one was created along the separatrix between the two flux systems of the quadrupole.
	
The emergence of sub-surface flux tubes into a simpler dipole coronal field configuration was simulated, in an attempt to remove  the twist field of the flux tube which first emerges into the corona. Reconnection at the separatrix between emerging field and pre-existing dipole field yielded an isolated flux rope, as did the deformation of the center of the emerging low tension flux tube

Although the coronal flux ropes that are formed in these simulations of flux emergence transport shear field into the low $\beta$ corona, the shear field is weak compared to the twist field and they are laden with photospheric/chromospheric  material. Any outward directed forces from the shear field are insufficient to overcome the inward
directed tension force and the downward directed gravity force. These coronal flux ropes are therefore
not candidates for erupting structures, and eventually disappear or fall back into the lower atmosphere. 

We have covered a range of flux tube profiles and coronal field structures and strengths, and in all cases found that the emergence of sub-surface flux tubes in these 2.5D simulations is an inefficient method for the transfer of magnetic shear energy into the low $\beta$ corona. The amount of magnetic shear energy supplied to the low $\beta$ corona was calculated, at heights of 1.2 Mm and 5 Mm. The lower of these two 
heights corresponds to the lowest point at which $\beta=0.2$ in the simulations in this paper, and thus the lowest point at which the conditions are the same as those used in CME models.
In all cases, the amount of magnetic shear energy supplied to the $\beta < 0.2$ 
corona was less than 10\% of the total magnetic energy and appears insufficient to directly drive a CME.
{\it We conclude that the simple observation of flux emergence and shear at the photosphere is not sufficient evidence that a CME can be driven.} Further investigations must be carried out to help understand the conditions under which emerging shear can be transported up to the low corona, where it is needed. 

For 3D flux tubes with significant curvature along the tube axis, flows along fieldlines may be an important phenomena in flux emergence simulations. These simulations should be repeated in 3D to see if this has an affect on the formation of coronal flux ropes and their eventual evolution in the solar atmosphere.

We have tested the robustness of these results by varying the resolution and introducing asymmetry into the flux emergence and coronal fields, and find similar results regardless of the position of the flux tube relative to the center of the quadrupole/dipole.

In these simulations, and almost all previous flux emergence simulations, the high $\beta$ lower atmosphere is assumed to be fully ionized, which is 
not the case for the solar chromosphere. The presence of neutrals in the nearly neutral region near the temperature minimum may have an important effect on these results. 
\citet{2006A&A...450..805L} and \citet{2007ApJ...666..541A} showed that this nearly neutral region can dramatically affect flux emergence. In particular, cross-field currents in emerging structures are destroyed by ion-neutral collisions, and
the magnetic field lines are able to slip through the neutral gas during emergence. Thus the flux tubes lift up less mass from the chromosphere when the effects of partial ionization are taken into account. All this occurs lower down in the solar atmosphere around the $\beta = 1$ region and 
will directly affect the transfer of both mass and shear energy to the low $\beta$ corona above. These effects, along with 3D effects,  may provide a mechanism for the transfer of magnetic shear energy into the corona without the transfer of mass, 
which may be key to driving a CME with dynamical flux emergence. This topic will be investigated in future publications.

\begin{acknowledgments}
J. E. Leake and M. G. Linton acknowledge support from NASA SR\&T grant number NNH06AD58I, from ONR/NRL 6.1
basic research funds,  and from the NRL-Hinode analysis program. Hinode is a Japanese mission developed
and launched by ISAS/JAXA, collaborating with NAOJ as domestic partner, and NASA (USA) and STFC (UK) 
as international  partners. Scientific operation of the Hinode mission is conducted by the Hinode science team
organized at ISAS/JAXA. This team mainly consists of scientists from institutes in the partner countries.
Support for the post-launch operation is provided by JAXA and NAOJ, STFC, NASA, ESA (European
Space Agency), and NSC (Norway). We are grateful to the Hinode team for all their efforts in the design,
build and operation of the mission. The work by S. K. Antiochos was supported by the NASA HTP, TR\&T, and SR\&T Programs.
The authors would like to thank C. R. DeVore for enlightening discussion concerning the magnetic breakout model for
CME initiation and the numerical modeling of CMEs. 

\end{acknowledgments}

\bibliography{leake_may_2010}

\begin{thebibliography}{}

\bibitem[{Acheson}, 1979]{1979SoPh...62...23A}
{Acheson}, D.~J. 1979,
\newblock \solphys, 62, 23

\bibitem[{Amari} et~al., 2007]{2007ApJ...671L.189A}
{Amari}, T., {Aly}, J.~J., {Mikic}, Z., \& {Linker}, J. 2007,
\newblock \apjl, 671, L189

\bibitem[{Amari} et~al., 2003]{2003ApJ...585.1073A}
{Amari}, T., {Luciani}, J.~F., {Aly}, J.~J., {Mikic}, Z., \& {Linker}, J. 2003,
\newblock \apj, 585, 1073

\bibitem[{Amari} et~al., 2000]{2000ApJ...529L..49A}
{Amari}, T., {Luciani}, J.~F., {Mikic}, Z., \& {Linker}, J. 2000,
\newblock \apjl, 529, L49

\bibitem[{Antiochos}, 1998]{1998ApJ...502L.181A}
{Antiochos}, S.~K. 1998,
\newblock \apjl, 502, L181+

\bibitem[{Antiochos} et~al., 1999]{1999ApJ...510..485A}
{Antiochos}, S.~K., {DeVore}, C.~R., \& {Klimchuk}, J.~A. 1999,
\newblock \apj, 510, 485

\bibitem[{Arber} et~al., 2007]{2007ApJ...666..541A}
{Arber}, T.~D., {Haynes}, M., \& {Leake}, J.~E. 2007,
\newblock \apj, 666, 541

\bibitem[{Arber} et~al., 2001]{2001JCoPh.171..151A}
{Arber}, T.~D., {Longbottom}, A.~W., {Gerrard}, C.~L., \& {Milne}, A.~M. 2001,
\newblock Journal of Computational Physics, 171, 151

\bibitem[{Arber} et~al., 1999]{1999ApJ...517..990A}
{Arber}, T.~D., {Longbottom}, A.~W., \& {van der Linden}, R.~A.~M. 1999,
\newblock \apj, 517, 990

\bibitem[{Archontis}, 2008]{2008JGRA..11303S04A}
{Archontis}, V. 2008,
\newblock Journal of Geophysical Research (Space Physics), 113, 3

\bibitem[{Archontis} et~al., 2006]{2006ApJ...645L.161A}
{Archontis}, V., {Galsgaard}, K., {Moreno-Insertis}, F., \& {Hood}, A.~W. 2006,
\newblock \apjl, 645, L161

\bibitem[{Archontis} \& {Hood}, 2008]{2008ApJ...674L.113A}
{Archontis}, V. \& {Hood}, A.~W. 2008,
\newblock \apjl, 674, L113

\bibitem[{Archontis} \& {Hood}, 2010]{2010A&A...514A..56A}
{Archontis}, V. \& {Hood}, A.~W. 2010,
\newblock \aap, 514, A56+

\bibitem[{Archontis} et~al., 2007]{2007A&A...466..367A}
{Archontis}, V., {Hood}, A.~W., \& {Brady}, C. 2007,
\newblock \aap, 466, 367

\bibitem[{Archontis} et~al., 2004]{2004A&A...426.1047A}
{Archontis}, V., {Moreno-Insertis}, F., {Galsgaard}, K., {Hood}, A., \&
  {O'Shea}, E. 2004,
\newblock \aap, 426, 1047

\bibitem[{Archontis} \& {T{\"o}r{\"o}k}, 2008]{2008A&A...492L..35A}
{Archontis}, V. \& {T{\"o}r{\"o}k}, T. 2008,
\newblock \aap, 492, L35

\bibitem[{Athay} \& {Thomas}, 1961]{1961psc..book.....A}
{Athay}, R.~G. \& {Thomas}, R.~N. 1961,
\newblock {Physics of the solar chromosphere},
\newblock (Interscience Monographs and Texts in Physics and Astronomy, New
  York: Interscience Publication, 1961)

\bibitem[{Bychenkov} et~al., 1988]{1988PhR...164..119B}
{Bychenkov}, V.~Y., {Silin}, V.~P., \& {Uryupin}, S.~A. 1988,
\newblock \physrep, 164, 119

\bibitem[{Chen} \& {Shibata}, 2000]{2000ApJ...545..524C}
{Chen}, P.~F. \& {Shibata}, K. 2000,
\newblock \apj, 545, 524

\bibitem[{DeVore} \& {Antiochos}, 2005]{2005ApJ...628.1031D}
{DeVore}, C.~R. \& {Antiochos}, S.~K. 2005,
\newblock \apj, 628, 1031

\bibitem[{DeVore} \& {Antiochos}, 2008]{2008ApJ...680..740D}
{DeVore}, C.~R. \& {Antiochos}, S.~K. 2008,
\newblock \apj, 680, 740

\bibitem[{Dubey} et~al., 2006]{2006A&A...459..927D}
{Dubey}, G., {van der Holst}, B., \& {Poedts}, S. 2006,
\newblock \aap, 459, 927

\bibitem[{Emonet} \& {Moreno-Insertis}, 1998]{1998ApJ...492..804E}
{Emonet}, T. \& {Moreno-Insertis}, F. 1998,
\newblock \apj, 492, 804

\bibitem[{Fan}, 2001]{2001ApJ...554L.111F}
{Fan}, Y. 2001,
\newblock \apjl, 554, L111

\bibitem[{Fan}, 2009]{2009ApJ...697.1529F}
{Fan}, Y. 2009,
\newblock \apj, 697, 1529

\bibitem[{Fan} \& {Gibson}, 2007]{2007ApJ...668.1232F}
{Fan}, Y. \& {Gibson}, S.~E. 2007,
\newblock \apj, 668, 1232

\bibitem[{Fisher} et~al., 2000]{2000SoPh..192..119F}
{Fisher}, G.~H., {Fan}, Y., {Longcope}, D.~W., {Linton}, M.~G., \& {Pevtsov},
  A.~A. 2000,
\newblock \solphys, 192, 119

\bibitem[{Forbes}, 2000]{2000JGR...10523153F}
{Forbes}, T.~G. 2000,
\newblock \jgr, 105, 23153

\bibitem[{Gary}, 2001]{2001SoPh..203...71G}
{Gary}, G.~A. 2001,
\newblock \solphys, 203, 71

\bibitem[{Gary} \& {Alexander}, 1999]{1999SoPh..186..123G}
{Gary}, G.~A. \& {Alexander}, D. 1999,
\newblock \solphys, 186, 123

\bibitem[{Gerrard} et~al., 2001]{2001A&A...373.1089G}
{Gerrard}, C.~L., {Arber}, T.~D., {Hood}, A.~W., \& {Van der Linden}, R.~A.~M.
  2001,
\newblock \aap, 373, 1089

\bibitem[{Gilman}, 1970]{1970ApJ...162.1019G}
{Gilman}, P.~A. 1970,
\newblock \apj, 162, 1019

\bibitem[{Gosling} et~al., 1976]{1976SoPh...48..389G}
{Gosling}, J.~T., {Hildner}, E., {MacQueen}, R.~M., {Munro}, R.~H., {Poland},
  A.~I., \& {Ross}, C.~L. 1976,
\newblock \solphys, 48, 389

\bibitem[{Hildner}, 1977]{1977ASSL...71....3H}
{Hildner}, E. 1977,
\newblock {Mass ejections from the solar corona into interplanetary space},
\newblock in Study of Travelling Interplanetary Phenomena,  ed. {M.~A.~Shea,
  D.~F.~Smart, \& S.~T.~Wu}, volume~71 of {\em Astrophysics and Space Science
  Library}, pages 3--20

\bibitem[{Karpen} et~al., 1996]{1996ApJ...460L..73K}
{Karpen}, J.~T., {Antiochos}, S.~K., \& {Devore}, C.~R. 1996,
\newblock \apjl, 460, L73+

\bibitem[{Kliem} \& {T{\"o}r{\"o}k}, 2006]{2006PhRvL..96y5002K}
{Kliem}, B. \& {T{\"o}r{\"o}k}, T. 2006,
\newblock Physical Review Letters, 96, 255002

\bibitem[Klimchuk, 2001]{Klimchuk2001}
Klimchuk, J.~A. 2001,
\newblock in Space Weather,  ed. .~H.~S. P.~Song, G.~Siscoe, volume Geophysical
  Monograph 125, (Washington: AGU),  142

\bibitem[{Leake} \& {Arber}, 2006]{2006A&A...450..805L}
{Leake}, J.~E. \& {Arber}, T.~D. 2006,
\newblock \aap, 450, 805

\bibitem[Linton \& Moldwin, 2009]{LintonM2009}
Linton, M.~G. \& Moldwin, M.~B. 2009,
\newblock JGR Space Physics, 114

\bibitem[{Lynch} et~al., 2008]{2008ApJ...683.1192L}
{Lynch}, B.~J., {Antiochos}, S.~K., {DeVore}, C.~R., {Luhmann}, J.~G., \&
  {Zurbuchen}, T.~H. 2008,
\newblock \apj, 683, 1192

\bibitem[{Lynch} et~al., 2004]{2004ApJ...617..589L}
{Lynch}, B.~J., {Antiochos}, S.~K., {MacNeice}, P.~J., {Zurbuchen}, T.~H., \&
  {Fisk}, L.~A. 2004,
\newblock \apj, 617, 589

\bibitem[{MacNeice} et~al., 2004]{2004ApJ...614.1028M}
{MacNeice}, P., {Antiochos}, S.~K., {Phillips}, A., {Spicer}, D.~S., {DeVore},
  C.~R., \& {Olson}, K. 2004,
\newblock \apj, 614, 1028

\bibitem[{MacTaggart} \& {Hood}, 2009a]{2009A&A...501..761M}
{MacTaggart}, D. \& {Hood}, A.~W. 2009a,
\newblock \aap, 501, 761

\bibitem[{MacTaggart} \& {Hood}, 2009b]{2009A&A...508..445M}
{MacTaggart}, D. \& {Hood}, A.~W. 2009b,
\newblock \aap, 508, 445

\bibitem[{Magara}, 2001]{2001ApJ...549..608M}
{Magara}, T. 2001,
\newblock \apj, 549, 608

\bibitem[{Manchester} et~al., 2004]{2004ApJ...610..588M}
{Manchester}, IV, W., {Gombosi}, T., {DeZeeuw}, D., \& {Fan}, Y. 2004,
\newblock \apj, 610, 588

\bibitem[{Murray} \& {Hood}, 2008]{2008A&A...479..567M}
{Murray}, M.~J. \& {Hood}, A.~W. 2008,
\newblock \aap, 479, 567

\bibitem[{Newcomb}, 1961]{1961PhFl....4..391N}
{Newcomb}, W.~A. 1961,
\newblock Physics of Fluids, 4, 391

\bibitem[{Parker}, 1955]{1955ApJ...122..293P}
{Parker}, E.~N. 1955,
\newblock \apj, 122, 293

\bibitem[{Parker}, 1988]{1988ApJ...326..407P}
{Parker}, E.~N. 1988,
\newblock \apj, 326, 407

\bibitem[Priest \& Forbes, 2000]{Priest2000}
Priest, E.~R. \& Forbes, T.~G. 2000,
\newblock Magnetic Reconnection: MHD Theory and Applications,
\newblock (New York: Cambridge University Press)

\bibitem[{Roussev} et~al., 2003]{2003ApJ...588L..45R}
{Roussev}, I.~I., {Forbes}, T.~G., {Gombosi}, T.~I., {Sokolov}, I.~V.,
  {DeZeeuw}, D.~L., \& {Birn}, J. 2003,
\newblock \apjl, 588, L45

\bibitem[{Schuessler}, 1979]{1979A&A....71...79S}
{Schuessler}, M. 1979,
\newblock \aap, 71, 79

\bibitem[{Shibata} et~al., 1989a]{1989ApJ...338..471S}
{Shibata}, K., {Tajima}, T., {Matsumoto}, R., {Horiuchi}, T., {Hanawa}, T.,
  {Rosner}, R., \& {Uchida}, Y. 1989a,
\newblock \apj, 338, 471

\bibitem[{Shibata} et~al., 1989b]{1989ApJ...345..584S}
{Shibata}, K., {Tajima}, T., {Steinolfson}, R.~S., \& {Matsumoto}, R. 1989b,
\newblock \apj, 345, 584

\bibitem[{Soenen} et~al., 2009]{2009A&A...501.1123S}
{Soenen}, A., {Zuccarello}, F.~P., {Jacobs}, C., {Poedts}, S., {Keppens}, R.,
  \& {van der Holst}, B. 2009,
\newblock \aap, 501, 1123

\bibitem[{Titov} \& {D{\'e}moulin}, 1999]{1999A&A...351..707T}
{Titov}, V.~S. \& {D{\'e}moulin}, P. 1999,
\newblock \aap, 351, 707

\bibitem[{van der Holst} et~al., 2007]{2007ApJ...671L..77V}
{van der Holst}, B., {Jacobs}, C., \& {Poedts}, S. 2007,
\newblock \apjl, 671, L77

\bibitem[{Vernazza} et~al., 1981]{1981ApJS...45..635V}
{Vernazza}, J.~E., {Avrett}, E.~H., \& {Loeser}, R. 1981,
\newblock \apjs, 45, 635

\bibitem[{Zhang} \& {Wu}, 2009]{2009JGRA..11405107Z}
{Zhang}, T.~X. \& {Wu}, S.~T. 2009,
\newblock Journal of Geophysical Research (Space Physics), 114, 5107

\bibitem[{Zirin}, 1970]{1970SoPh...14..328Z}
{Zirin}, H. 1970,
\newblock \solphys, 14, 328

\bibitem[{Zuccarello} et~al., 2008]{2008ApJ...689L.157Z}
{Zuccarello}, F.~P., {Soenen}, A., {Poedts}, S., {Zuccarello}, F., \& {Jacobs},
  C. 2008,
\newblock \apjl, 689, L157

\bibitem[{Zwaan}, 1987]{1987ARA&A..25...83Z}
{Zwaan}, C. 1987,
\newblock \araa, 25, 83

\end{thebibliography}

\eject

\clearpage

\end{document}